\documentclass{aa}
\usepackage[varg]{txfonts}
\usepackage{natbib}
\usepackage{hyperref}
\usepackage{booktabs}
\usepackage{amsmath}
\usepackage{multirow}

\bibpunct{(}{)}{;}{a}{}{,} 
\usepackage{graphicx}
\usepackage{xcolor}
\definecolor{myblue}{rgb}{0 0.349 0.702}

\hypersetup{
	colorlinks=true,
	linkcolor=myblue,
    citecolor=myblue,
	urlcolor=myblue,
	citebordercolor=myblue,
	linkbordercolor=myblue,
	urlbordercolor=white
	}
\makeatletter
\renewcommand*\aa@pageof{, page \thepage{} of \pageref*{LastPage}}
\makeatother

\titlerunning{Exoplanet Characterization using cINNs}
\title{Exoplanet Characterization using Conditional Invertible Neural Networks}
\author{Jonas Haldemann\inst{1,2} 
	\and Victor Ksoll\inst{3} 
	\and Daniel Walter\inst{3}
	\and Yann Alibert\inst{1} 
	\and Ralf S.\ Klessen \inst{3,4}
	\and Willy Benz \inst{1}
	\and Ullrich Koethe \inst{5}
	\and Lynton Ardizzone \inst{5}
	\and Carsten Rother \inst{5}}
\date{\today}

\offprints{Jonas Haldemann, \email{jonas.haldemann@unibe.ch}}

\institute{Departement of Space Research \& Planetary Sciences, University of Bern, Gesellschaftsstrasse 6, 3012 Bern, Switzerland
\and Abteilung Physik, Gymnasium Lerbermatt, Kirchstrasse 64, 3098 Köniz, Switzerland
\and
Universität Heidelberg, Zentrum für Astronomie, Institut für Theoretische Astrophysik, Albert-Ueberle-Straße 2, 69120 Heidelberg, Germany
\and
Universität Heidelberg, Interdisziplinäres Zentrum für Wissenschaftliches Rechnen, Im Neuenheimer Feld 205, D-69120 Heidelberg, Germany
\and
Computer Vision and Learning Lab (HCI, IWR), Universität Heidelberg Berliner Str. 43 D-69120 Heidelberg, Germany
}

\date{Received 31 January 2022 / Accepted ??}

\abstract{The characterization of an exoplanet‘s interior is an inverse problem, which requires statistical methods such as Bayesian inference in order to be solved. Current methods employ Markov Chain Monte Carlo (MCMC) sampling to infer the posterior probability of planetary structure parameters for a given exoplanet. These methods are time consuming since they require the evaluation of a planetary structure model $\sim 10^5$ times.} {To speed up the inference process when characterizing an exoplanet, we propose to use conditional invertible neural networks to calculate the posterior probability of the planetary structure parameters. } {Conditional invertible neural networks (cINNs) are a special type of neural network which excel in solving inverse problems. We constructed a cINN following the framework for easily invertible architectures (FreIA). This neural network was then trained on a database of $5.6 \cdot 10^6$ internal structure models to recover the inverse mapping between internal structure parameters and observable features (i.e., planetary mass, planetary radius and elemental composition of the host star). We show in this work also how one can account for observational uncertainties.} {The cINN method was compared to a commonly used Metropolis-Hastings MCMC. For that we repeated the characterization of the exoplanet K2-111 b, using both the MCMC method and  the trained cINN. We show that the inferred posterior probability distributions of the internal structure parameters from both methods are very similar, with the biggest differences seen in the exoplanet's water content. Thus cINNs are a possible alternative to the standard time-consuming sampling methods. Indeed, using cINNs allows for orders of magnitude faster inference of an exoplanet's composition than what is possible using an MCMC method, however, it still requires the computation of a large database of internal structures to train the neural network. Since this database is only computed once, we found that using an invertible neural network is more efficient than an MCMC, when more than 10 exoplanets are characterized using the same neural network.} {}
\keywords{planetary systems -- planets and satellites: interiors -- methods: numerical -- methods: data analysis}
\begin{document}

\maketitle

\section{Introduction}\label{Sec:Introduction}

More than a decade ago exoplanetary science has entered the era of characterization, where new observations are used to infer physical and chemical properties of exoplanets. These properties can be related to the atmosphere \citep[e.g.,][]{hoeijmakers_spectral_2019,madhusudhan_exoplanetary_2019} or planetary composition \citep[e.g.,][]{Dorn2015}. In the latter case, mass and radius measurements of an exoplanet are used to derive its internal structure (e.g., size of the iron core, presence of water, gas mass fraction, etc.). This problem is notoriously strongly degenerate \citep[][]{rogers_framework_2010}, though part of this degeneracy can be removed when assuming that the bulk refractory composition of an exoplanet matches the one of its parent star \citep{dorn_generalized_2017}. This assumption is supported by numerical simulations \citep[e.g.,][]{thiabaud_elemental_2015} as well as solar system observations \citep[e.g.,][]{sotin_massradius_2007}, although studies of observed exoplanets are not yet conclusive \citep{plotnykov_chemical_2020,schulze_probability_2020,adibekyan_chemical_2021}

Even in this latter case the problem of deriving the planetary composition from mass, radius and refractory composition remains degenerate. The traditional method is to use Bayesian inference where the posterior probability of planetary structure parameters is derived from the set of observed parameters, given prior probability distributions on the planetary structure parameters. Such Bayesian calculations are in general performed using a Markov Chain Monte Carlo (MCMC) method \citep[see][]{mosegaard_monte_1995,Dorn2015,dorn_generalized_2017,haldemann_mcmc_2021}.

MCMC methods are an efficient and well tested way of sampling probability distributions. 
One does not need an analytical description of the whole normalized probability density function (PDF) of the target distribution. Instead one only needs to be able to calculate ratios of the PDF at pairs of locations in the phase space \citep{hogg_data_2018}. This means that in the case of Bayesian inference, one only needs to be able to compute the product of the prior probability and the likelihood function, skipping the expensive calculation of Bayes' integral, which acts as a normalization constant. 

Although using MCMC sampling together with a planetary structure model has led to a number of successful planetary characterizations \citep[e.g.,][]{dorn_bayesian_2017,agol_refining_2020}, it suffers from at least two major difficulties. The first one is that when planetary parameters are measured with small error bars\footnote{This is the case for radius determination using high precision telescopes like CHEOPS \citep{benz_cheops_2017,benz_cheops_2021} or PLATO in the future \citep{rauer_space_2018}.}, the likelihood function becomes very narrow which, depending on the used MCMC method, can drastically increase the needed time to converge to a solution. The second difficulty lies in the consideration of multi-planetary systems. When characterizing multiple exoplanets simultaneously, a planetary structure model needs to be calculated for every considered exoplanet, which linearly increases the computational cost. At the same time, the number of parameters that characterize the planetary compositions in a multi-planetary system is similarly scaled with the number of exoplanets. This increase of the dimensionality of the parameter space also implies generally an increase in the number of points needed to properly sample the posterior distribution, and therefore an increase in the required computing time, which is overall more than linear.

In this paper, we propose a new method to derive the posterior distribution of planetary structure parameters. This method is based on conditional invertible neural networks (cINNs), a type of neural network architecture that is able to provide the posterior distribution of planetary structure parameters for any given choice of observed parameters (e.g., mass, radius, refractory composition of the host star). 
One key aspect of the proposed method is that the distribution of planetary structure parameters predicted by the cINN matches in principle (in the limit of an infinitely accurate neural network - see Sect. \ref{Sec:Methods}) any given set of observed parameters, for which the network was trained, {\it without any measurement error}. As a consequence, the cINN does not suffer from the first difficulty we mentioned above, as it is naturally suited for the case of high-precision measurements with very small uncertainties. 
Another key aspect is that, once trained, the cINN provides the posterior distribution of planetary structure parameters in a few minutes, where modern MCMCs often require hours or even days to converge due to the time consuming evaluation of the forward model (see Sect. \ref{Sec:Results}).
cINNs have also already seen successful applications in astronomy. \citet{ksoll_stellar_2020} have managed to estimate stellar parameters from photometric observations of resolved star clusters using a cINN and \citet{kang_emission-line_2022} recently showed a cINN approach to recover physical parameters of star-forming clouds from spectral observations.

This paper is structured in the following way. In Sect. \ref{Sec:Methods} we describe the basic concept of invertible neural networks and conditional invertible neural networks in particular. We show how they can be set up in order to characterize exoplanets and how the forward model works that generates the training data for the neural networks. In Sect. \ref{Sec:method_val} we first validate our proposed method using a simple toy model. Then in Sect. \ref{Sec:Results} we apply the proposed method to characterize an observed exoplanet and compare its performance to a regular Metropolis-Hastings MCMC, which was previously used for the same purpose. In Sect. \ref{Sec:Discussion} we discuss the current limitations of the approach and compare the time required to run either an MCMC or use the proposed method for cINNs. Finally, we summarize our findings in Sect. \ref{Sec:Conclusions}.

\section{Methods}\label{Sec:Methods}
\subsection{Invertible Neural Networks}\label{Sec:INN_Methods}
\begin{figure}
    \centering
    \includegraphics[width = \linewidth]{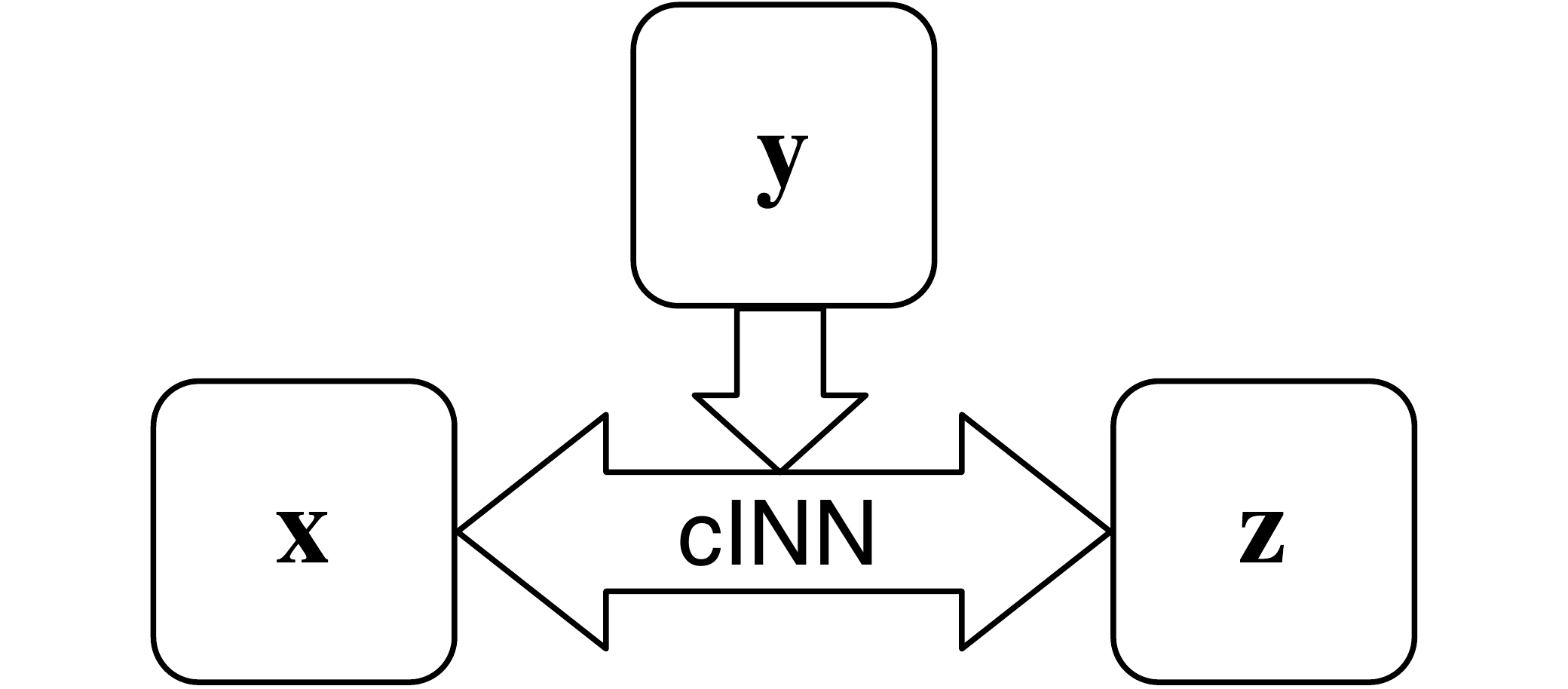}
    \caption{Schematic overview over the cINN. During training the cINN learns to encode all information about the physical parameters $\mathbf{x}$ in the latent variables $\mathbf{z}$ (while enforcing that these follow a Gaussian distribution) that is not contained in the observations $\mathbf{y}$. At prediction time, conditioned on the new observation $\mathbf{y}$, the cINN then transforms the known prior distribution $p(\mathbf{z})$ to $\mathbf{x}$-space to retrieve posterior distribution $p(\mathbf{x}|\mathbf{y})$.}
    \label{fig:cINN_schematic}
\end{figure}

The invertible neural network (INN) provides an architecture that excels in solving inverse problems \citep{Ardizzone2019a}. In these problems one often has access to a well understood forward model (e.g., a simulation) that describes the mapping between, e.g., underlying physical parameters $\mathbf{x}$ of an object and their corresponding observable quantities $\mathbf{y}$. At the same time, however, recovering the inverse mapping $\mathbf{y} \rightarrow \mathbf{x}$, that is of central interest in many applications, is a difficult task. The INN approach to these inverse problems makes the additional assumption, that the known forward model always induces some form of information loss in the mapping $\mathbf{x} \rightarrow \mathbf{y}$, which can be encoded in some unobservable, latent variables $\mathbf{z}$. Leveraging a fully invertible architecture the INN is then trained to approximate the known forward model $f$, learning to associate $\mathbf{x}$ values with unique pairs of $[\mathbf{y}, \mathbf{z}]$, i.e., a bijective mapping. In doing so, it automatically provides a solution for the inverse mapping $f^{-1}$ for free. For simplicity  \citep[as described in][]{Ardizzone2019a} it is further assumed that the latent variables $\mathbf{z}$ follow a Gaussian prior distribution, which is enforced during the training process. Note, however, in principle any desired distribution can be prescribed for the latent priors. 

Given a new observation $\mathbf{y}$ this procedure allows to predict the full posterior distribution $p(\mathbf{x}|\mathbf{y})$ by simply sampling from the known prior distribution of the latent space. The architecture of the INN consists of a series of reversible blocks following a design proposed by \citet{Dinh2016}. After splitting the input vector $\mathbf{u}$ into two halves $\mathbf{u_1},\,\mathbf{u_2}$ these blocks perform two complementary affine transformations 
\begin{align}
    \label{eq:INN_forward1}
            \mathbf{v_1} &= \mathbf{u_1} \odot \exp\left(s_2\left(\mathbf{u_2}\right)\right) + t_2\left(\mathbf{u_2}\right),\\
    \label{eq:INN_forward2}
            \mathbf{v_2} &= \mathbf{u_2} \odot \exp\left(s_1\left(\mathbf{v_1}\right)\right) + t_1\left(\mathbf{v_1}\right),
\end{align}
using element-wise multiplication $\odot$ and addition. Here $s_i$ and $t_i$ denote arbitrarily complex mappings of $\mathbf{u_2}$ and $\mathbf{v_1}$,  e.g., like small fully connected networks, which are not required to be invertible as they are only ever evaluated in the forward direction. 

\noindent Inverting these affine transformations is trivial following
\begin{align}
    \label{eq:INN_backward1}
        \mathbf{u_2} &= \left(\mathbf{v_2} - t_1\left(\mathbf{v_1}\right)\right) \odot \exp\left(-s_1\left(\mathbf{v_1}\right)\right),\\
    \label{eq:INN_backward2}
        \mathbf{u_1} &= \left(\mathbf{v_1} - t_2\left(\mathbf{u_2}\right)\right) \odot \exp\left(-s_2\left(\mathbf{u_2}\right)\right).
\end{align}
\subsubsection{Conditional Invertible Neural Networks}\label{Sec:CINN_Methods}
In this paper we employ a modification to this approach called conditional invertible neural network (cINN) as proposed in \cite{Ardizzone2019b} and previously applied in \citet{ksoll_stellar_2020}. Here the affine coupling blocks are adapted to accept a conditioning input $\mathbf{c}$ such that the mappings in Eqs.~\eqref{eq:INN_forward1} - \eqref{eq:INN_backward2}, i.e., $s_2(\mathbf{u_2})$,  $t_2(\mathbf{u_2})$, etc., are replaced with  $s_2([\mathbf{u_2}, \mathbf{c}])$ and $t_2([\mathbf{u_2}, \mathbf{c}])$, respectively. By concatenating conditions to the inputs of the subnetworks like this, the invertibility of the architecture is not affected. Note that both the forward $f(\mathbf{x};\mathbf{c}) = \mathbf{z}$ and backward mapping $\mathbf{x} = g(\mathbf{z}; \mathbf{c})$ of the cINN entail this conditioning, and that the invertibility of the network is given for \textit{fixed} condition $\mathbf{c}$ as $f(\cdot\,;\mathbf{c})^{-1} = g(\cdot\,; \mathbf{c})$. 

When using cINNs for inverse regression problems, like an exoplanet's internal structure characterization, the observations $\mathbf{y}$ (e.g., planetary mass, radius and stellar refractory composition) serve as the conditioning input. Figure~\ref{fig:cINN_schematic} shows a schematic representation of the cINN in this case. In doing so the cINN, just like the INN, will learn to encode all information about the physical parameters $\mathbf{x}$ into the latent variables $\mathbf{z}$ that is not contained in the observations $\mathbf{y}$ during training. Aside from usually delivering better results, the cINN approach has the additional advantage that no zero padding is needed if the dimension of $[\mathbf{y}, \mathbf{z}]$ is larger than the dimension of $\mathbf{x}$, as we can simply set $\dim(\mathbf{z}) = \dim(\mathbf{x})$ \citep{Ardizzone2019a,Ardizzone2019b}. 

Given the condition $\mathbf{c}$ of a new observation $\mathbf{y}$ the posterior distribution of the physical parameters is, as for the INN, determined by sampling the latent variables $\mathbf{z}$ from their Gaussian prior
\begin{equation}
    p(\mathbf{x}|\mathbf{y}) = g(\mathbf{z}; \mathbf{c} = \mathbf{y})\,\,\mathrm{with}\, z \sim p_Z(\mathbf{z}) = \mathbb{N}(\mathbf{z}, 0, \mathbf{I}), \label{Eq:posterior_cinn}
\end{equation}
where $\mathbf{I}$ is the $K\times K$ unity matrix with $K = \mathrm{dim}(\mathbf{z})$.
In this framework, prior information on $\textbf{x}$ is learned by the network from the distribution of $\textbf{x}$ in the set of training data. That means that the distribution of the training data should follow the prior probability distribution $p(\textbf{x})$.

\subsubsection{Network Architecture}
\begin{figure}
    \centering
    \resizebox{1.\hsize}{!}{\includegraphics{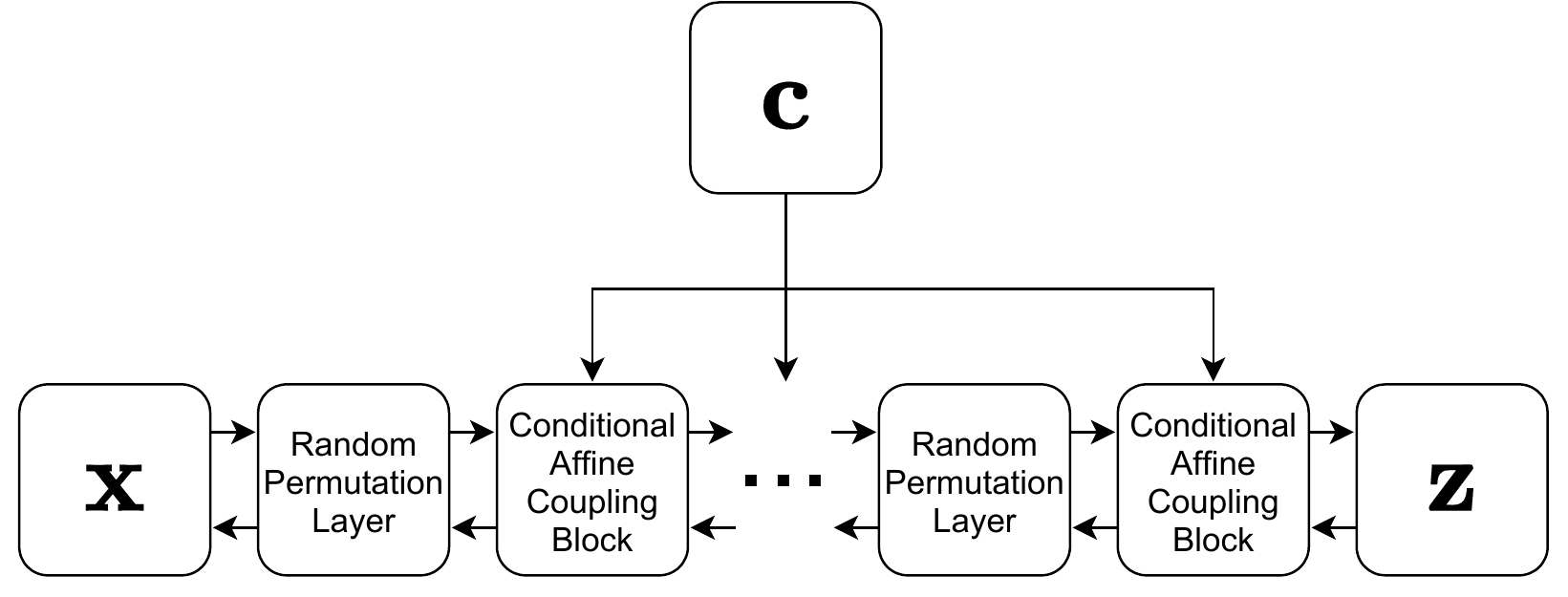}}
    \caption{Schematic overview over the cINN architecture.}
    \label{fig:cINN_architecture_overview}
\end{figure}

For the work presented in this paper we employ the ’Framework for Easily Invertible Architectures’ \citep[FrEIA,][]{Ardizzone2019a, Ardizzone2019b} and mostly follow the specific cINN architecture suggested in \cite{Ardizzone2019b}. This means we alternate reversible blocks in the GLOW configuration \citep{Kingma2018} with random permutation layers (see Fig. \ref{fig:cINN_architecture_overview} for a schematic of the structure). The former is a computationally efficient variant, where the outputs of the mappings $s_i(\cdot)$ and $t_i(\cdot)$ are predicted jointly by a single subnetwork instead of one each. We use simple three layer (width of 512 per layer) fully connected networks with rectified linear unit (ReLU) activation functions for these subnetworks. The random permutation layers use random orthogonal matrices, which are fixed during training and cheaply invertible, to better mix the information between the streams $\mathbf{u_1}$ and $\mathbf{u_2}$. Together with the structure of the affine transformations this ensures that the cINN cannot just ignore the conditioning input during training. In total our architecture consists of 8 reversible blocks. Note that contrary to \cite{Ardizzone2019b} we do not apply a feature extraction network that transforms the input conditions into an intermediate representation because of the low dimensionality of the observable parameter space in our problem. Some early experiments have shown that such a network did not benefit the predictive performance of the cINN on the given regression task. We train the cINN by minimization of the maximum likelihood loss. We refer to \cite{Ardizzone2019b} for further details on this matter.\\

\subsubsection{Data Pre-Processing}
Before any training and prediction is performed, the data is pre-processed. First we transform both physical parameters $\mathbf{x}$ and observables $\mathbf{y}$ into log-space. This ensures that the physical parameters stay strictly positive, while it also reduces magnitude differences between different physical quantities. This is important since vastly different magnitudes between parameters can cause training instabilities, e.g., as a single parameter could dominate the target function (loss) that is minimized during the training procedure. To further address this issue we also center both the physical parameters $\mathbf{x_i}$ and observables $\mathbf{y_i}$, and re-scale them, such that their standard deviations become unity. These \textit{linear} scaling transformations are easily inverted at prediction time. 
Note that the scaling transformation parameters are derived from the training set and that the \textit{same} transformations are applied to new data at prediction time. 

\subsubsection{Evaluating Training Performance}
To quantify the success of the cINN training procedure we proceed as in \citet{ksoll_stellar_2020}. We  measure its performance on a test data set, a held-out subset of the training data consisting of 20,000 randomly selected synthetic observations. On this test set we then first confirm whether the distribution of latent variables has converged to match the target multivariate normal distribution with unit covariance matrix.

Afterwards, we evaluate the shape of the predicted posterior distributions by computing the median calibration error $s$, as proposed in \cite{Ardizzone2019a}, for each of the target parameters $x$. Given an uncertainty interval $q$ the calibration error $e_\mathrm{cal, q}$ for a collection of $N$ observations is defined as the difference
\begin{equation}
    e_\mathrm{cal, q} = q_\mathrm{inliers} - q,
\end{equation}
where $q_\mathrm{inliers} = N_\mathrm{inliers}/N$ denotes the fraction of observations, where the true value $\tilde{x}$ lies within the $q$-confidence interval of the predicted posterior PDF. Values of $e_\mathrm{cal,q} < 0$ signify that the predicted PDFs are too narrow, whereas positive values suggest the opposite, i.e. that the PDFs are too broad. The median calibration error $s$ is derived as the median of the absolute calibration errors over the range of confidences from 0 to 1.\\
Next we quantify the cINNs predictive capability for maximum a posteriori (MAP) point estimates $\hat{x}$. To do so we derive an accuracy for the individual target parameters $x$ over the entire test set as given by the root mean squared error (RMSE) and normalized RMSE (NRMSE). They are defined as
\begin{equation}
    \mathrm{RMSE} = \sqrt{\frac{\sum_{i=1}^{N} \left(\hat{x}_i - \tilde{x}_i\right)^2}{N}},
\end{equation}
where $\tilde{x}_i$ is the ground truth value of the target parameter for the $i$-th observation, and
\begin{equation}
    \mathrm{NRMSE} = \frac{\mathrm{RMSE}}{\bar{x}},
\end{equation}
where $\bar{x} = x_\mathrm{max}^\mathrm{ts} -x_\mathrm{min}^\mathrm{ts}$ denotes the range of parameter $x$ within the training data. To determine the MAP estimates $\hat{x}$ from the predicted samples of the posterior distribution, we perform a kernel density estimation (KDE) to model the PDF and find its maximum. This KDE employs a Gaussian kernel function and is computed on an evenly spaced grid of $1024$ points, covering the full range of the given posterior samples. The kernel bandwidth $h$ is derived using Silverman's rule \citep{Silverman1986}, 
\begin{equation}
    h = 1.06 \cdot \min\left(\sigma, \frac{\mathrm{IQR}}{1.34}\right) \cdot n^{-\frac{1}{5}},
\end{equation}
where $\mathrm{IQR}$, $\sigma$ and $n$ denote the interquartile range, standard deviation and number of the posterior samples, respectively.

\subsubsection{Predicting Posteriors for Noisy Observations}\label{Sec:posterior_with_uncert}
\begin{figure}
    \centering
    \includegraphics[width = \linewidth]{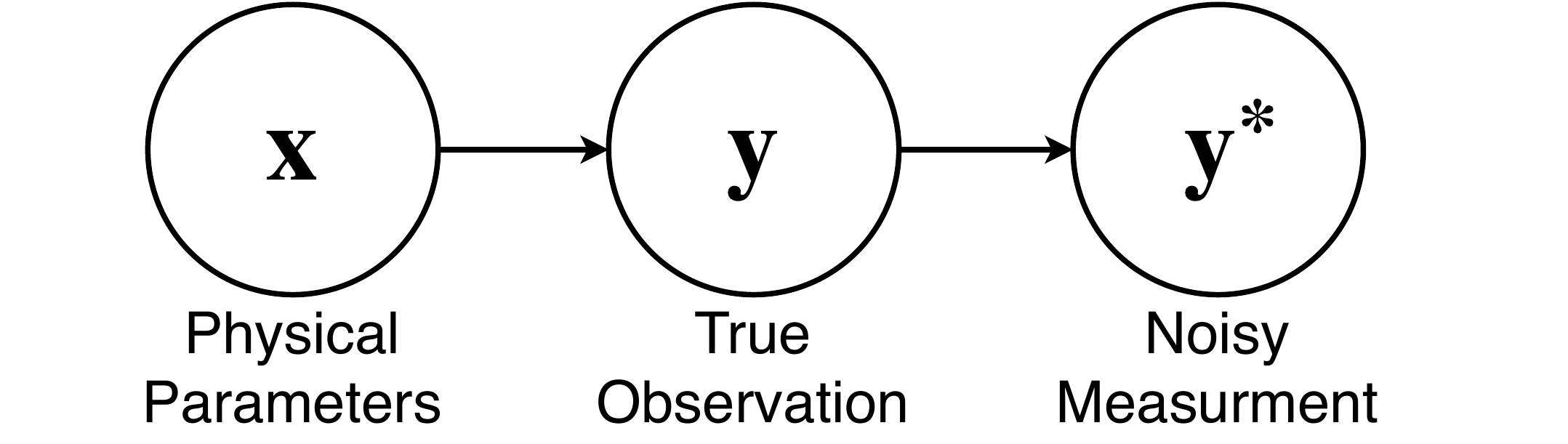}
    \caption{Simple graph representation for the problem of predicting physical parameters from noisy measurements of a true observable quantity.}
    \label{fig:cINN_Graph_noisy.pdf}
\end{figure}

As the method was outlined so far, the cINN does not include the possibility that a given input observation can be uncertain. Instead the described method assumed perfect observations as an input. However, in many real world applications all observed quantities usually suffer from measurement uncertainties. In order to predict posterior probability distribution of $\textbf{x}$ given a noisy observation using the cINN, we devise the following strategy.

Let the noisy observation be represented by $\textbf{y}^\ast$ and the true observable properties of the target be denoted as $\textbf{y}$ (as shown in the graph representation in Fig.~\ref{fig:cINN_Graph_noisy.pdf}). For this paper we assume that the distribution of $\textbf{y}^\ast$ follows a multivariate normal distribution of dimension $k$ with mean $\boldsymbol{\mu}$ and covariance $\boldsymbol{\Sigma}$, i.e.,
\begin{equation}
\textbf{y}^\ast\sim\mathbb{N}_k(\boldsymbol{\mu},\boldsymbol{\Sigma}). \label{Eq:noisy_observ}
\end{equation}
Given the law of total probability, the posterior probability distribution $p(\textbf{x} \mid \textbf{y}^\ast)$ can be written as \begin{equation}
	p(\textbf{x} \mid \textbf{y}^\ast) = \int_Y p(\textbf{x} \mid \textbf{y}^\ast \cap \textbf{y}=\textbf{y}^\prime)\Phi_{\textbf{y}^\prime\mid \textbf{y}^\ast}(\textbf{y}^\prime)\, d\textbf{y}^\prime\;,
\end{equation}
where $\textbf{y}^\prime$ is a point in the space of observational parameters and $\Phi_{\textbf{y}^\prime\mid \textbf{y}^\ast}$ is the probability density function of $\textbf{y}^\prime$ given $\textbf{y}^\ast$. Since we assume that $\textbf{y}^\ast$ follows a multivariate normal distribution, $\Phi_{\textbf{y}^\prime\mid \textbf{y}^\ast}$ is given by
\begin{equation}
\Phi_{\textbf{y}^\prime\mid \textbf{y}^\ast}(\textbf{y}^\prime) = \frac{1}{(2\pi)^{k/2}\sqrt{\det(\boldsymbol{\Sigma})}}\exp\left(-\frac{1}{2}(\textbf{y}^\prime-\boldsymbol{\mu})^T\boldsymbol{\Sigma}^{-1}(\textbf{y}^\prime-\boldsymbol{\mu})\right).
\end{equation}
Next we use that \textbf{x} is conditionally independent of $\textbf{y}^\ast$ given $\textbf{y}$, i.e., $((\textbf{x}\perp\!\!\!\perp \textbf{y}^\ast)\mid\, \textbf{y})$. It follows that $p(\textbf{x}\mid\textbf{y}^\ast\cap \textbf{y}) = p(\textbf{x}\mid \textbf{y})$ and  hence
\begin{equation}
p(\textbf{x} \mid \textbf{y}^\ast) = \int_Y p(\textbf{x} \mid \textbf{y}=\textbf{y}^\prime)\Phi_{\textbf{y}^\prime\mid \textbf{y}^\ast}(\textbf{y}^\prime)\, d\textbf{y}^\prime,
\end{equation}
where $p(\textbf{x}\mid \textbf{y}=\textbf{y}^\prime)$ can be calculated for a given $\textbf{y}^\prime$ from Eq.~\eqref{Eq:posterior_cinn}.
The posterior probability distribution can now be calculated using simple Monte Carlo integration:
\begin{itemize}
	\item The Monte Carlo samples of $\textbf{x}$ are generated by first drawing $N$ times a sample $\textbf{y}^\prime_i$ from the multivariate normal distribution given in Eq.~\eqref{Eq:noisy_observ}.
	\item For each sample $\textbf{y}^\prime_i$ one calculates the point estimated of  $p(\textbf{x}| \textbf{y}^\prime_i)$ using the cINN as outlined in  Sec.~\ref{Sec:CINN_Methods}.
	For each $\textbf{y}^\prime_i$ one therefore samples another $M$ times from the latent variables $\textbf{z}$ and evaluates for each $\textbf{z}_i$ the backward mapping of the cINN, i.e., $g(\textbf{z}_i,\textbf{c}=\textbf{y}^\prime_i)$.
\end{itemize}    
This results in $N$ $\times$ $M$ samples of $\textbf{x}$ drawn from the posterior probability distribution $p(\textbf{x}\mid\textbf{y}^\ast)$. 

Note that by definition the conditional probability $p(\textbf{x}\mid\textbf{y})$ is zero if $p(\textbf{x})=0$ or $p(\textbf{y})=0$.
Hence if the prior distribution of $\textbf{x}$ has a compact support, then $p(\textbf{x}\mid \textbf{y})$ is automatically zero for any $\textbf{x}$ outside of the domain of $\textbf{x}$.
While the extent of the domain is in principle learned by the cINN, it is still possible that the cINN will map some $\textbf{z}_i$ to an $\textbf{x}$ for which $p(\textbf{x})=0$.
During the sampling of $\textbf{z}$ all such samples should thus be rejected.
Additionally, the compact support of $p(\textbf{x})$ will simultaneously limit the possible output values of the forward model and therefore also induce limits on $\textbf{y}$.
One can hence forgo to evaluate the cINN for any $\textbf{y}^\prime_i$ for which $p(\textbf{y}^\prime_i)=0$.
The kind of limits introduced for $\textbf{y}$ depend on the  forward model.
An aspect which we discuss in more detail in Sec.~\ref{Sec:limits} .

\subsection{Forward Model}\label{Sec:forward}
\begin{figure}
    \centering
	\resizebox{0.9\hsize}{!}{\includegraphics{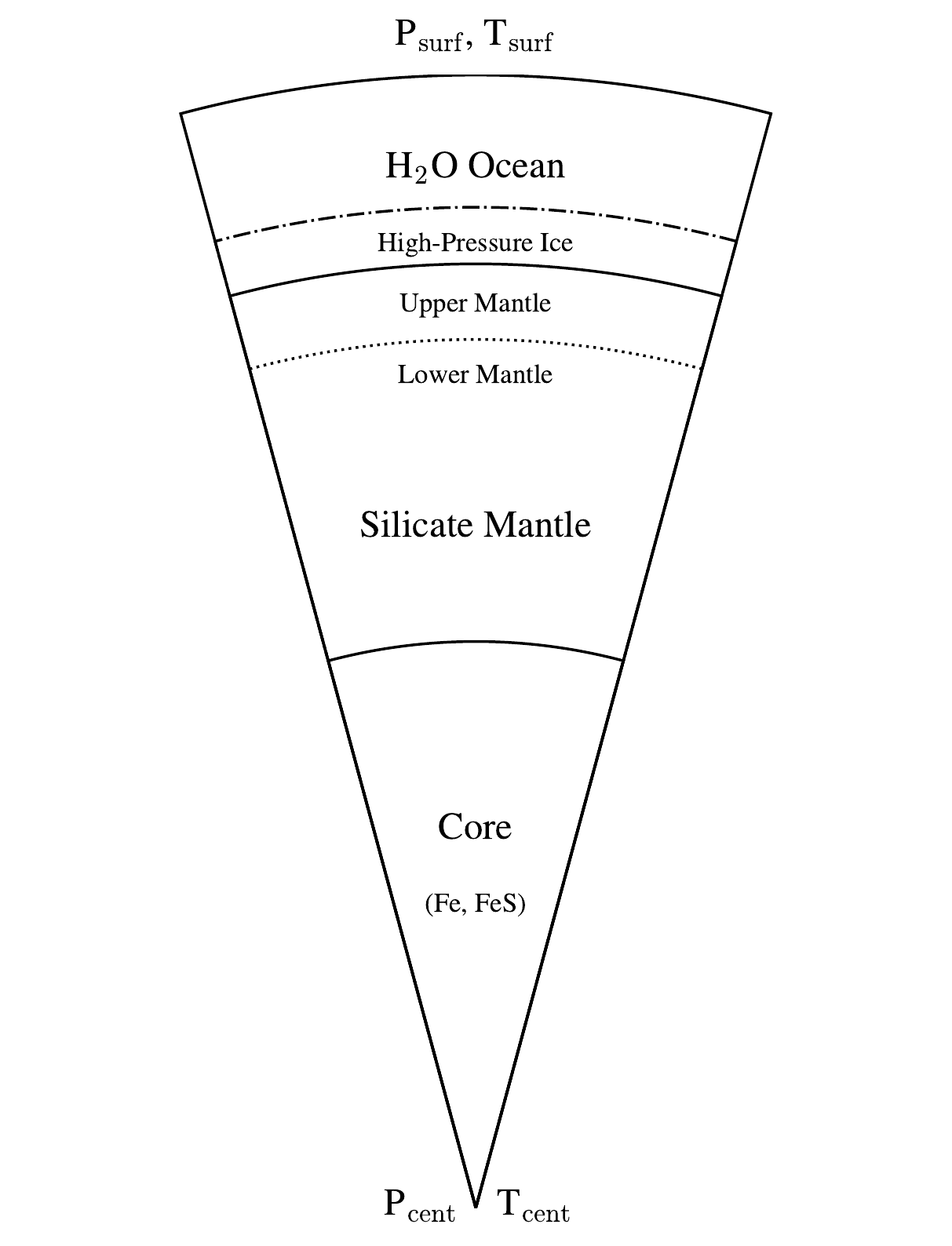}}
	\caption{Schematic representation of the layered planetary structure. Three main layers are present, the core, silicate mantle and volatile layer. Depending on the size of the layers an upper mantle can be present if the volatile layer above is not too massive. Contrary if the volatile layer is massive enough high pressure ices might form on the bottom of the layer. }
	\label{Fig:planet_schematic}
\end{figure}

The forward model $f$ maps the physical input parameters $\mathbf{x}$ to a prediction in the data space $\textbf{y}^\prime$, i.e.,
\begin{equation}
    f(\textbf{x}) = \textbf{y}^\prime.
\end{equation}
It does so by calculating the interior structure of a 1D-spherically symmetric sphere in hydrostatic equilibrium. Following \citet{kippenhahn_stellar_2012} and similar to the case of stellar structures, we solve the two point boundary value problem given by the equations 
\begin{align}
\frac{\partial r}{\partial m} &= \frac{1}{4\pi r^2 \rho},\label{EQ:mass_cons} \\
\frac{\partial P}{\partial m} &= -\frac{Gm}{4\pi r^4}, \label{EQ:hydrostatic_eq}\\
\frac{\partial T}{\partial m} &= \frac{\partial P}{\partial m }\frac{T}{P}\nabla, \label{EQ:thermal_transport}
\end{align}
where $r$ is the radius, $m$ the mass within radius $r$, $P$ the pressure, $T$ the temperature, $\rho$ the density, $G$ the gravitational constant and $\nabla$ is the dimensionless temperature gradient. The sphere is split into three layers of distinct composition akin to a differentiated planet (see Fig. \ref{Fig:planet_schematic}). The thermodynamic properties in each layer are given by a set of equations of state (EoS) listed in Table \ref{Tab:eos_table}.

From the EoS we calculate $\rho$, the thermal expansion coefficient $\alpha$ and the specific heat capacity $c_P$. We assume that each layer is in a regime of vigorous convection. Therefore, the dimensionless temperature gradient is given by the adiabatic temperature gradient
\begin{equation}
\nabla = \left(\frac{\partial \ln{T}}{\partial \ln{P}}\right)_S= \frac{\alpha P}{\rho c_p}.
\end{equation}

\begin{table}
	\caption{List of equation of state (EoS) used in the forward model.}
	\label{Tab:eos_table}
	\centering
	\begin{tabular}{c c c}
		\hline\hline
		Layer & Composition & EoS \\
		\hline
		\multirow{2}*{Core} & \multirow{2}*{Fe, S} & \citet{hakim_new_2018},\\ &&\citet{fei_thermal_2016}\tablefootmark{a} \\
		Mantle & Fe, Mg, Si, O&  \citet{sotin_massradius_2007}\\
		Volatile & H$_2$O& \citet{haldemann_aqua_2020} \\
		\hline
		
		\hline
	\end{tabular}
		\tablefoot{
	\tablefoottext{a}{Only for pressures below 310 GPa.}
		}
\end{table}

\subsubsection{Core}
We consider a solid iron core made out of hcp-Fe with possible inclusions of less dense FeS alloys. In the model, the composition of the core is given by the  sulphur fraction $\left. x_\text{S}\right|_\text{Core}$, i.e.,
\begin{align}
	\left. x_\text{Fe}\right|_\text{Core} &= \frac{1-2\left. x_\text{S}\right|_\text{Core}}{1-\left. x_\text{S}\right|_\text{Core}},\\
	\left.x_\text{FeS}\right|_\text{Core} &= \frac{\left. x_\text{S}\right|_\text{Core}}{1-\left. x_\text{S}\right|_\text{Core}}. 
\end{align}
 The thermodynamic properties of Fe and FeS within the core were calculated using the EoS of \citet{hakim_new_2018}. However for Fe at  pressures below 310 GPa the EoS of \citet{fei_thermal_2016} was used, as it is advised in the work of \citet{hakim_new_2018}.

\subsubsection{Mantle}
The mantle structure was calculated following the model used in \citet{sotin_massradius_2007}. It assumes a homogeneous elemental composition of Fe, Mg, Si and O throughout the mantle considering four different minerals.
In the upper mantle the model includes the iron and magnesium end members of the minerals \textit{olivine} ([Mg,Fe]$_2$SiO$_4$) and \textit{ortho pyroxene} ([Mg,Fe]$_2$Si$_2$O$_6$), while for the lower mantle a composition of \textit{perovskite} ([Mg,Fe]SiO$_3$) and \textit{wüstite} ([Mg,Fe]O) is assumed.
The respective fractions of the mineral phases are calculated from the ratios of the $x_\text{Mg}/x_\text{Si}|_\text{Mantle}$ and $x_\text{Fe}/x_\text{Si}|_\text{Mantle}$ mole fractions as in \citet{sotin_massradius_2007}, where
\begin{equation}
\left.\frac{x_\text{Mg}}{x_\text{Si}}\right|_\text{Mantle} = \frac{\left.x_\text{MgO} \right|_\text{Mantle}}{\left.x_\text{SiO$_2$} \right|_\text{Mantle}}
\end{equation}
and
\begin{equation}
\left.\frac{x_\text{Fe}}{x_\text{Si}}\right|_\text{Mantle} = \frac{\left.x_\text{FeO} \right|_\text{Mantle}}{\left.x_\text{SiO$_2$} \right|_\text{Mantle}}.
\end{equation}
Note that due to the assumption of a homogeneous elemental composition and the choice of minerals in the model of \citet{sotin_massradius_2007}, the  $x_\text{Mg}/x_\text{Si}|_\text{Mantle}$ and $x_\text{Fe}/x_\text{Si}|_\text{Mantle}$ ratios are limited by the possible spread in said minerals.  Thus only compositions which fulfil the relation
\begin{equation}
1\leq \left.\frac{x_\text{Mg}}{x_\text{Si}}\right|_\text{Mantle}+\left.\frac{x_\text{Fe}}{x_\text{Si}}\right|_\text{Mantle}  \leq 2 \label{Eq:sotin_condition}
\end{equation}
can be calculated with this model.
This ultimately also limits the possible Mg to Fe and Si to Fe ratios of the whole exoplanet.
The resulting limits will be further discussed in Sec. \ref{Sec:limits}.

\subsubsection{Volatiles}
The outermost volatile layer is assumed to be entirely made up of H$_2$O. We forwent including an additional H/He layer in order to reduce the number of model parameters and hence the time needed to calculate the database of forward models used to train the cINN. Though more realistic volatile layers are planned to be added in the future. The EoS of H$_2$O is given by the AQUA-EoS of \citet{haldemann_aqua_2020}  which combines the \textit{ab initio} EoS of \citet{mazevet_2019} with the EoS of the high pressure ices (VII and X) by \citet{french_redmer_2015}, the EoS for ice II-VI by \citet{journaux_holistic_2020}, the EoS for ice Ih by \citet{feistel_new_2006} and the EoSs by \citet{wagner_iapws_2002}, \citet{brown_local_2018}, \citet{cea1_1994} and \citet{cea2_1996} for the liquid and vapor regions where \citet{mazevet_2019} is not applicable. 

\subsubsection{Numerical Method}
To solve the two point boundary value problem of Eqs. (\ref{EQ:mass_cons})-(\ref{EQ:thermal_transport}), we use a so-called bidirectional shooting method. 
That means given the set of input parameters listed in Table \ref{Tab:forward_params}, the Eqs. (\ref{EQ:mass_cons})-(\ref{EQ:thermal_transport}) are integrated using a 5th order Cash-Karp Runge-Kutta method with adaptive step size control \citep{nr_press}. This integration yields as output the total radius of the planet. The two remaining output variables, the planet's Mg/Fe and Si/Fe ratios, can be calculated from the core and mantle composition and the respective layer mass fractions.
\begin{table}
	\caption{List of forward model parameters.}
	\label{Tab:forward_params}
	\centering
	\begin{tabular}{@{}l r }
		\toprule
		Symbol & Parameter  \\
		\midrule
		\multicolumn{2}{c}{Input}\\
		\midrule
		M$_\text{tot}$ & the exoplanet's total mass\\
		$w_\text{core}$ & core mass fraction\\
		$w_\text{vol}$ & volatile mass fraction\\
		$x_\text{SiO$_2$}|_\text{mantle}$ & molar fraction of SiO$_2$ in the mantle\\
		$ x_\text{MgO}|_\text{mantle}$ & molar fraction of MgO in the mantle\\
		$x_\text{S}|_\text{core}$ & molar fraction of S in the core\\
		\midrule
		\multicolumn{2}{c}{Constants}\\
		\midrule
		T$_\text{surf} = 300$ K & surface temperature\\
		P$_\text{surf} = 1$ atm & surface pressure\\
		\midrule
		\multicolumn{2}{c}{Output}\\
		\midrule
		R$_\text{tot}$ & the exoplanet's total radius\\
		$x_\text{Mg}/x_\text{Fe}|_\text{Planet}$ & planetary ratio of Mg to Fe mole fractions\\
		$x_\text{Si}/x_\text{Fe}|_\text{Planet}$ & planetary ratio of Si to Fe mole fractions\\
		R$_\text{core}$ & the exoplanet's iron core radius\\
		R$_\text{mantle}$ & the exoplanet's mantle radius\\
		R$_\text{vol}$ & the exoplanet's volatile layer radius\\
		\bottomrule
	\end{tabular}
\end{table}

\subsection{Forward Model Limits}\label{Sec:limits}
As mentioned in Sec.~\ref{Sec:posterior_with_uncert} one has to consider in this method the limits of
the forward model and restrict the sampling of observable features $\textbf{y}$ to the domain of the training set. Otherwise, we find that if an observation is close to the domain boundary of the training set then the quality of the sampling suffers greatly since the cINN could not properly learn the inverse mapping for these regions.

From the used planetary structure model two sets of limits can be constructed for the parameters in the space of observable features. 

\subsubsection{Limits of Mg/Fe and Si/Fe}
The mantle model of \citet{sotin_massradius_2007} allows only for mantle compositions which fulfil Eq.~\eqref{Eq:sotin_condition}.
This range in possible mantle compositions can be translated to a limit for the possible bulk composition of the modelled exoplanets. The limits for the bulk composition are derived in the following way. In our structure model iron can occur both in the core and the mantle, whereas Mg and Si are only included in the mantle. Thus the Mg to Si ratio of the mantle always represents the Mg to Si ratio of the whole exoplanet.
The upper limit of the Mg to Si ratio, therefore, occurs when there is no iron in the mantle, i.e., when 
\begin{equation}
\left.\frac{x_\text{Mg}}{x_\text{Si}}\right|_\text{Mantle} =\left.\frac{x_\text{Mg}}{x_\text{Si}}\right|_\text{Planet} =2. \label{Eq:cinn_upper_mgsi0}
\end{equation}
Multiplying Eq.~\eqref{Eq:cinn_upper_mgsi0} with $x_\text{Si}/x_\text{Fe}|_\text{Planet}$ returns the upper limit on the planetary Mg to Fe ratio 
\begin{equation}
\left.\frac{x_\text{Mg}}{x_\text{Fe}}\right|_\text{Planet} = 2\left.\frac{x_\text{Si}}{x_\text{Fe}}\right|_\text{Planet}. \label{Eq:upper_bound1}
\end{equation}
The lower bound of the $x_\text{Mg}/x_\text{Si}|_\text{Planet}$ occurs when the maximum possible amount of iron is in the mantle, i.e., in exoplanets without a core where $x_\text{Fe}/x_\text{Si}|_\text{Mantle} = x_\text{Fe}/x_\text{Si}|_\text{Planet}$ . In that case one can write similarly to Eq.~\eqref{Eq:sotin_condition}
\begin{equation}
1\leq \left.\frac{x_\text{Mg}}{x_\text{Si}}\right|_\text{Planet}+ \left.\frac{x_\text{Fe}}{x_\text{Si}}\right|_\text{Planet}
\end{equation}
or in terms of iron abundance ratios
\begin{equation}
\left.\frac{x_\text{Si}}{x_\text{Fe}}\right|_\text{Planet}\leq \left.\frac{x_\text{Mg}}{x_\text{Fe}}\right|_\text{Planet}+ 1. \label{Eq:lower_bound1}
\end{equation}
Values of $x_\text{Si}/x_\text{Fe}|_\text{Planet}$ and $x_\text{Mg}/x_\text{Fe}|_\text{Planet}$ which do not fulfil Eqs.~\eqref{Eq:upper_bound1} and \eqref{Eq:lower_bound1} can not be modelled by the forward model, thus the prior probability of any such value is zero. This does not mean that in nature these values can not occur, it is simply a limitation of the current model.

\subsubsection{Limits of \texorpdfstring{M$_\text{tot}$}{Mtot} and \texorpdfstring{R$_\text{tot}$}{Rtot}}
Similar to the compositional output parameters one can determine the limits of the forward model for M$_\text{tot}$ and R$_\text{tot}$. The limiting relations are in this case given by the mass radius relation of the most and least dense composition. The densest composition of this forward model is given by a pure iron sphere. The corresponding mass radius relation is
\begin{equation}
\frac{R_\text{tot}}{R_\text{E}} = 0.796\cdot \left(\frac{M_\text{tot}}{M_\text{E}}\right)^{0.2485}. \label{Eq:lower_bound2}
\end{equation}
In contrast, the least dense composition we consider is an exoplanet consisting of 70 wt.\% of water and 30 wt.\% mantle, with a composition given by $x_\text{Mg}/x_\text{Si}|_\text{Mantle} = 2$ and $x_\text{Fe}/x_\text{Si}|_\text{Mantle} = 0$.
The corresponding mass radius relation is then
\begin{equation}
\frac{R_\text{tot}}{R_\text{E}} = 1.341\cdot\left( \frac{M_\text{tot}}{M_\text{E}}\right)^{0.2564}. \label{Eq:upper_bound2}
\end{equation}
Regarding the limits of the refractory elements any combinations of total mass and total radius which does not fulfil Eqs.~\eqref{Eq:lower_bound2} and \eqref{Eq:upper_bound2} can not be modelled with the used forward model.
We will see in Fig.~\ref{Fig:corner_training2} in  Sec.~\ref{Sec:app_training}, that these relations indeed bracket the generated training data of this forward model.

\subsection{Generation of the training data}\label{Sec:app_training}
\begin{figure}
    \centering
	\resizebox{0.9\hsize}{!}{\includegraphics{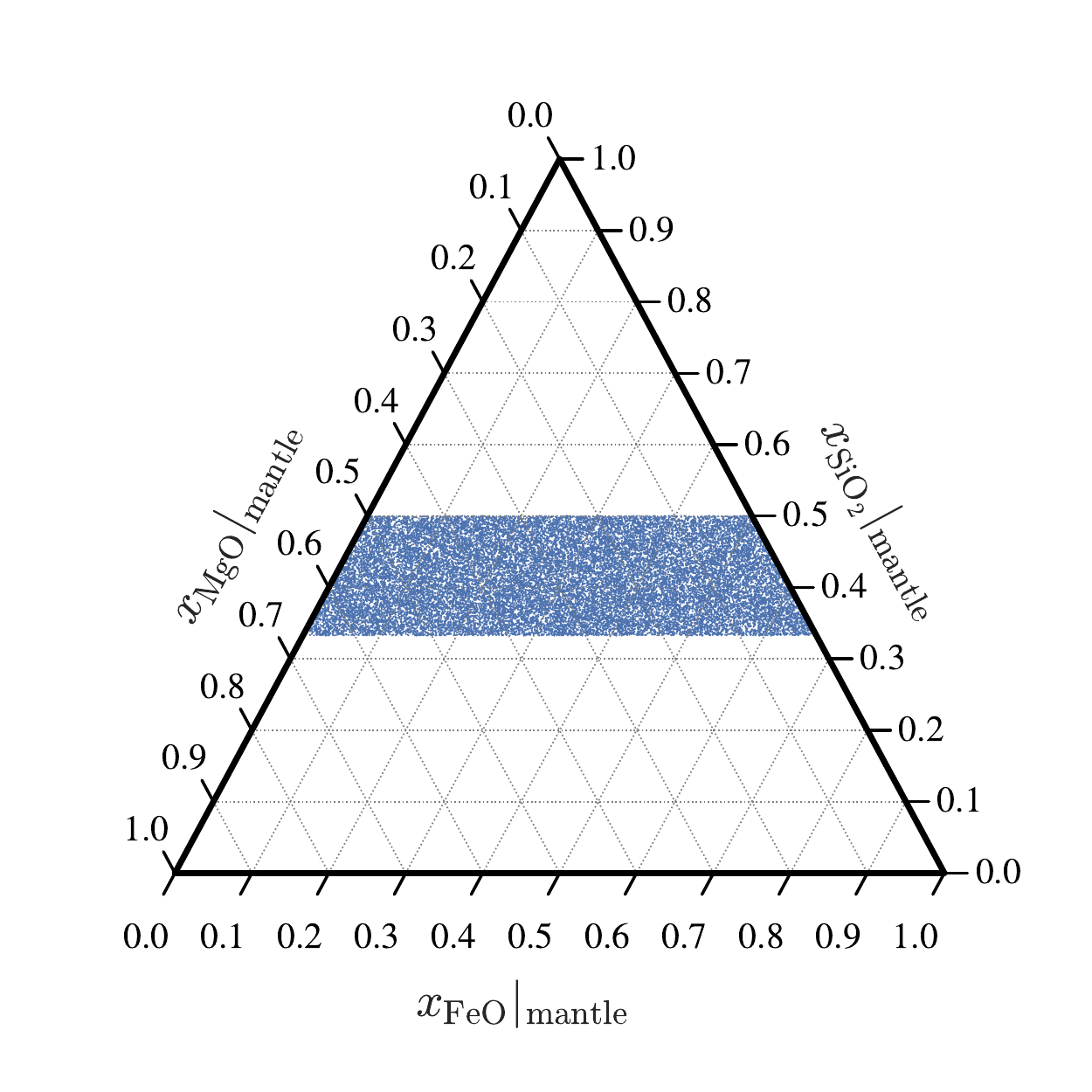}}
	\caption{Distribution of the SiO$_2$, MgO and FeO mole fractions of the training set.}
	\label{Fig:ternary_comp}
\end{figure}

\begin{figure}
    \centering
	\resizebox{1.05\hsize}{!}{\includegraphics{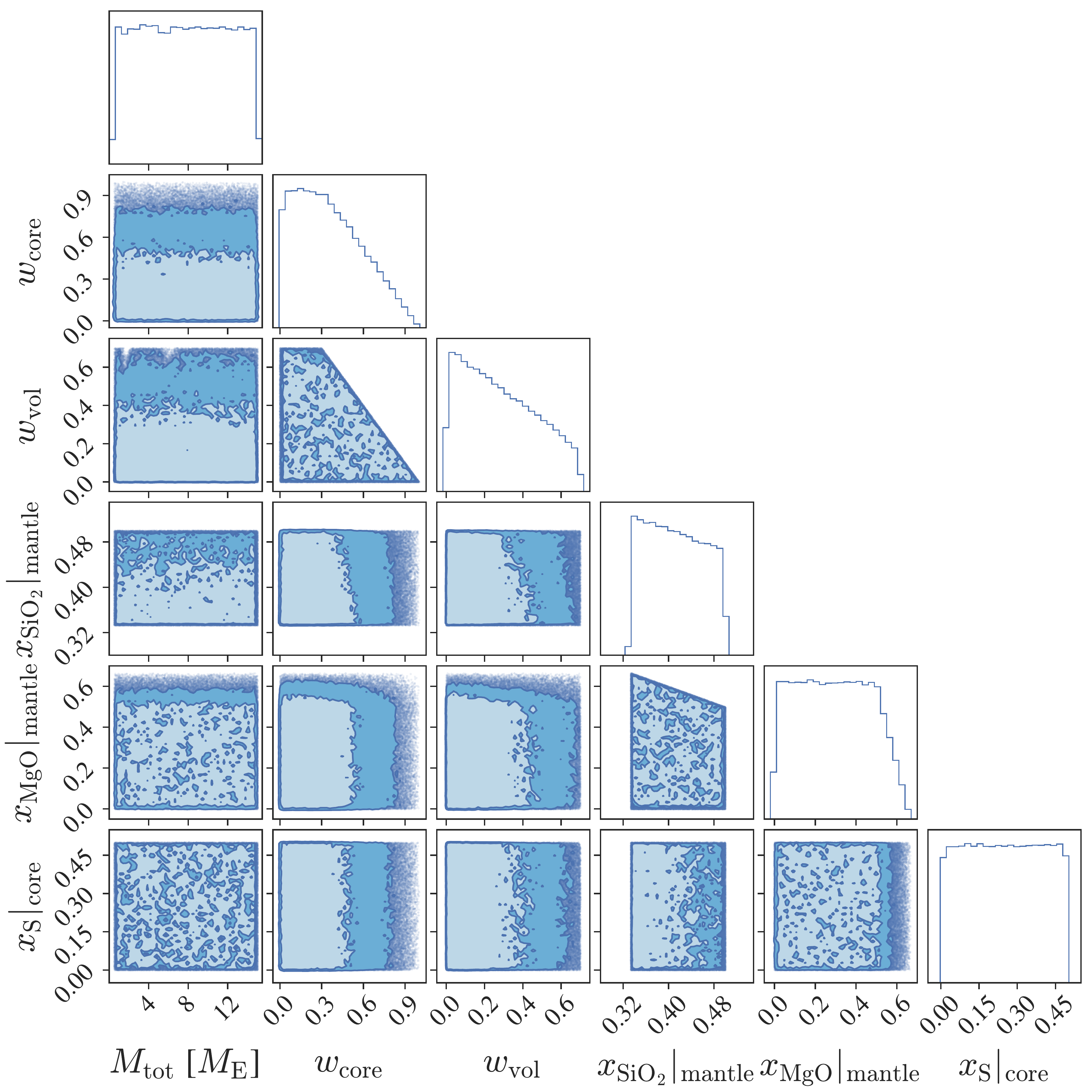}}
	\caption{Distribution of the forward model input parameters as generated for the training set. The underlying distributions from which the parameters were generated are listed in Table \ref{tab:input_dist}.}
	\label{Fig:corner_training1}
\end{figure}

\begin{figure}
	\centering
	\resizebox{\hsize}{!}{\includegraphics{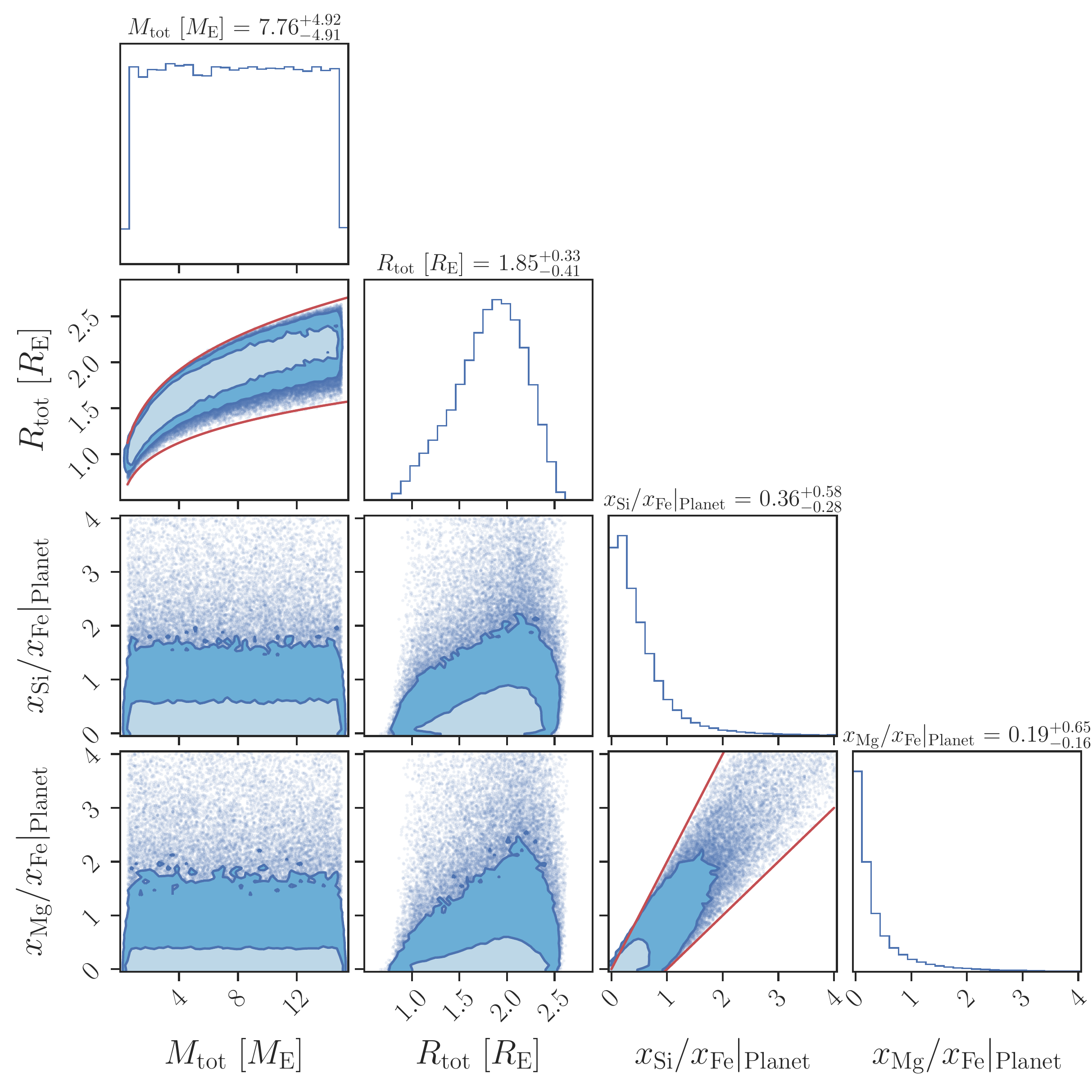}}
	\caption[Distribution of the forward model output parameters as generated for the training set]{Distribution of the forward model output parameters as generated for the training set. The underlying distributions from which the parameters were generated are listed in Table \ref{tab:input_dist}. The solid red lines indicate the limits of the forward model as described in Sect. \ref{Sec:limits}. The light shaded areas in the 2D diagrams indicate the 68\%-highest density region (HDR), while the dark shaded areas are the 89\%-HDR.}
	\label{Fig:corner_training2}
\end{figure}

In order to train the cINN we computed $5.9\cdot10^6$ forward models. The forward model input parameters were drawn at random from the distributions summarized in Table \ref{tab:input_dist}. The total mass of the planet was drawn from the uniform distribution $\mathbb{U}(0.5\text{ M}_\text{E},15 \text{ M}_\text{E})$. Since the layer mass fractions ($w_\text{core}$, $w_\text{mantle}$, $w_\text{vol}$) sum up to one per definition, they were drawn uniformly from the 3-dimensional probability simplex, with the restriction that the maximum water mass fraction cannot exceed a value of $0.7$. 
The mantle Si/Fe and Mg/Fe ratios were calculated from the mantle SiO$_2$, MgO and FeO mole fractions, which are the assumed sole constituents of the mantle model of \citet{sotin_massradius_2007} and hence sum up to one. Similar to the layer mass fractions we draw the SiO$_2$, MgO and FeO mole fractions uniformly from the 3d probability simplex. But since we use the mantle model of \citet{sotin_massradius_2007} an additional rejection sampling was performed excluding all combinations which did not fulfill Eq. (\ref{Eq:sotin_condition}). The resulting distribution on the probability simplex is shown in Fig. \ref{Fig:ternary_comp}.
The accepted values were then used to calculate the mantle Si/Fe and Mg/Fe ratios which are the forward model input parameters.
In the forward model the core is made of a mixture of Fe and FeS. Hence, we drew $\left.x_\text{S}\right|_\text{Core}$ from $\mathbb{U}(0,0.5)$ and used this value to calculate the S to Fe molar ratio in the core using
\begin{equation}
    \left.\frac{\text{S}}{\text{Fe}}\right|_\text{Core} = \frac{\left.x_\text{S}\right|_\text{Core}}{(1-\left.x_\text{S}\right|_\text{Core})}.
\end{equation}
The resulting distribution of all input and output parameters within the training set is shown in Figs. \ref{Fig:corner_training1} and \ref{Fig:corner_training2}.

\begin{table}[ht]
    \centering
    \caption{Distribution of the forward model input parameters within the training set.}
    \label{tab:input_dist}
    \begin{tabular}{@{}ll}
         \hline
         \hline
         Parameters & Distribution in training set\\
         \midrule
         $M_\text{tot}$ & $M_\text{tot}\sim \mathbb{U}(0.5 \text{ M}_\text{E},15\text{ M}_\text{E})$\\
		$w_\text{core}$, $w_\text{vol}$ & Uniform on $\Delta^2$, $w_\text{vol} = \min(w_\text{vol}, 0.7)$\\
		$\left. x_\text{SiO$_2$}\right|_\text{Mantle}$, $\left. x_\text{MgO}\right|_\text{Mantle}$ & Uniform on $\Delta^2$ and Eq.~\eqref{Eq:sotin_condition}\\
        $\left.x_\text{S}\right|_\text{Core}$ &   $\left.x_\text{S}\right|_\text{Core}\sim\mathbb{U}(0,0.5)$ \\ 
        \bottomrule         
    \end{tabular}
    \tablefoot{Here $\Delta^2$ denotes the 3 dimensional probability simplex.}
\end{table}


\section{Method Validation}\label{Sec:method_val}
In order to validate the sampling scheme outlined in Sect. \ref{Sec:Methods}, we use a simple toy model to benchmark the proposed scheme against a common Metropolis-Hastings MCMC sampler. 
We want to test here in particular how the sampling performs when: 

\begin{itemize}
    \item[i)] the posterior distribution of the model parameters has non zero probability along the boundary of the prior domain,
    \item[ii)] the observation is close to a region in which the forward model can not be applied anymore
\end{itemize}  

This will in particular demonstrate how the method performs, when an observation is close to the border of the set of training data.

To mimic the situations (i) and (ii) we setup the following toy model. We defined the space of model parameters $\mathbf{x} = (x_1,x_2) \in \mathbb{R}^2$, as well as the space of observable features $\mathbf{y}= (y_1,y_2) \in \mathbb{R}^2$. The forward model $f(\mathbf{x})$ is given by the linear model
\begin{equation}
    f\left(\begin{bmatrix}x_1\\ x_2\end{bmatrix}\right) = \begin{bmatrix}1 & 0\\ 1&2\end{bmatrix} \begin{bmatrix}x_1\\ x_2\end{bmatrix} = \begin{bmatrix}y_1\\ y_2\end{bmatrix}.
    \label{Eq:dummy_model}
\end{equation}
The prior domain is given by an equilateral triangle in the space of model parameters, defined by its corners 
\begin{equation}
    \mathbf{c}_1 = \begin{bmatrix}0 \\ 0\end{bmatrix},\, \mathbf{c}_2=\begin{bmatrix}1 \\ 0\end{bmatrix},\, \mathbf{c}_3=\begin{bmatrix}0.5 \\ 0.5\sqrt{3}\end{bmatrix}.
\end{equation}
For simplicity we choose a uniform prior probability distribution within the prior domain, thus p(\textbf{x}) is written as
\begin{equation}
    p(\textbf{x})=\begin{cases}
    \frac{4}{\sqrt{3}} & \mbox{if $\mathbf{x}$ within triangle($\mathbf{c}_1,\mathbf{c}_2,\mathbf{c}_3$)} \\
    0, & \mbox{if $\mathbf{x}$ outside triangle($\mathbf{c}_1,\mathbf{c}_2,\mathbf{c}_3$)} 
    \end{cases}
\end{equation}
where the probability to be within the triangle is the inverse of the area of the triangle. 
For a noisy observation $\textbf{y}^\ast$ defined as in Eq. \eqref{Eq:noisy_observ}, i.e., as a multivariate normal distribution
\begin{equation}
\textbf{y}^\ast \sim \mathbb{N}_2\left(\boldsymbol{\mu}=\begin{bmatrix}0.4\\1.3\end{bmatrix},\boldsymbol{\Sigma}=\begin{bmatrix}0.1^2 & 0 \\ 0 & 0.1^2 \end{bmatrix}\right),
\label{Eq:dummy_obs}
\end{equation}
the posterior distribution on \textbf{x} will become heavily truncated by the prior domain.
To mimic case (ii) we further added a restriction on the observable parameter $y_2$ and arbitrarily set an upper limit of
\begin{equation}
y_2 \leq 1.4. \label{Eq:cinn_dummy_lim}
\end{equation}

\begin{figure}
	\centering
	\resizebox{0.9\hsize}{!}{\includegraphics{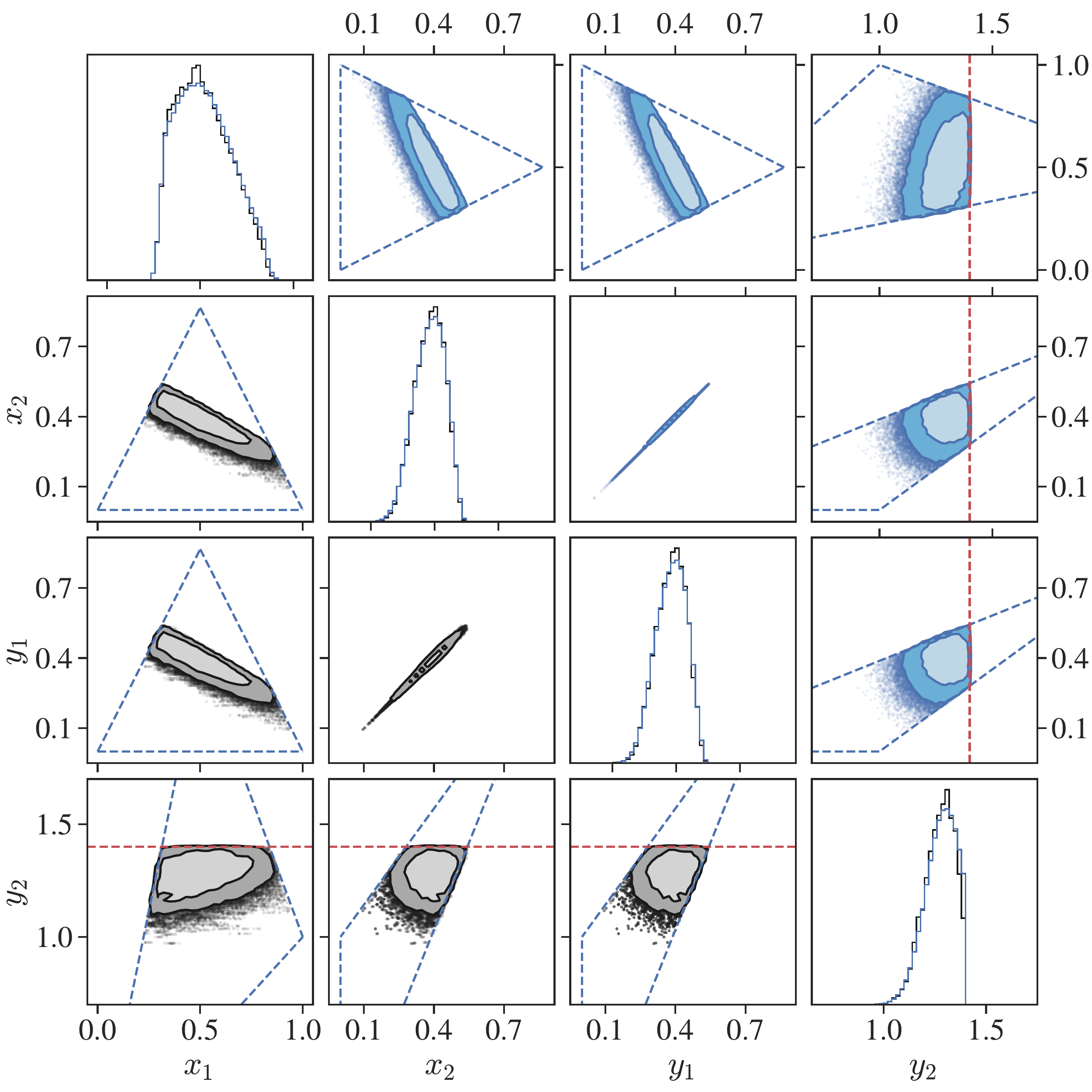}}
	\caption[Comparison of the cINN and a MCMC when sampling noisy data]{Comparison of the cINN and an MCMC method when applied to the toy model. 
	The data in the lower triangle (black) was predicted by the cINN method while the data in the upper triangle (blue) was generated with an MCMC sampler. The dashed blue lines indicate the boundaries of the prior domain, outside of which the prior probability is zero. The red dashed line indicates the upper limit on $y_2$ as in Eq.~\eqref{Eq:cinn_dummy_lim}. The histograms of the analytical solution are omitted since they overlap with the MCMC data. The light shaded areas in the 2D scatter plots indicate the 68\%-HDR, while the dark shaded areas are the 89\%-HDR.}
	\label{Fig:dummy_model}
\end{figure}

The cINN  was then trained on a dataset containing $10^6$ samples of $\textbf{x}$ and corresponding $f(\textbf{x})$ values.
Given $\textbf{y}^\ast$ as in Eq.~\eqref{Eq:dummy_obs} we then followed the method outlined in Sect.~\ref{Sec:posterior_with_uncert}, to sample from the posterior distribution.
The resulting distributions from the cINN and the MCMC sampling are shown in Fig.~\ref{Fig:dummy_model}, with the summary statistics of the marginalized distributions listed in Tab.~\ref{tab:dummy_stat}.
Looking at the summary statistics of the marginalized distributions, one can see that the median of each variable does not vary between methods.
Also the centered 1-$\sigma$ interval (containing 68.3\% of all samples) and the centered 2-$\sigma$ interval (containing 95.4\% of all samples) are almost identical (except for a 0.01 deviation of the lower bound of the 1-$\sigma$ interval of $y_2$ and the upper bound of the 2-$\sigma$ interval of $x_1$). To compare the shape of the resulting distributions we also computed the Hellinger distance $h$ between the marginalized probability distributions of the two methods.

The Hellinger distance $h(r,q)$ between two discrete probability distributions $r$ and $q$ is given by 
\begin{equation}
h(r,q)=\sqrt{1-b(r,q)},\label{Eq:hellinger1}
\end{equation}
where $b(r,q)$ is the Bhattacharyya coefficient \citep[see][]{hellinger_neue_1909},
\begin{equation}
b(r,q)=\sum_x\sqrt{r(x)q(x)} \label{Eq:hellinger2}.
\end{equation}
The Hellinger distance is a proper distance metric, it is 0 if the distributions $r$ and $q$ are identical and 1 if they are disjoint.  Here the Hellinger distance is determined from the histograms of the 1D marginalized distributions generated using the cINN or MCMC method.

We report that the Hellinger distance between the distributions for $x_1$ is $h=0.026$ whereas for $x_2$ it is $h=0.021$. For the observable features the Hellinger distance has similarly small values of $h=0.024$ for $y_1$ and $h=0.027$ for $y_2$. Such small values of the Hellinger distance are equivalent to the $h$ of two normal distributions with $\sigma=1$, where the median differs by $\sim 10^{-5}$ between the two normal distributions.

Looking at the 2D marginalized posterior densities in Fig.~\ref{Fig:dummy_model}, one can see that the posterior distribution on the input parameters $x_1$ and $x_2$ is strongly truncated, as it was expected.
Also the upper limit of $y_2$ has a notable effect on the distribution of the input parameters. 
But one does not see major differences between the two methods. We conclude that the cINN method can be successfully used for such a simple model, even when the observation is close to the  boundary of the training data.

\begin{table}
	\centering
	\caption[Summary statistics of the marginalized posterior distributions of model A]{Summary statistics (i.e., median, centered 1-$\sigma$ interval and centered 2-$\sigma$ interval) of the marginalized posterior distributions of the toy model}
	\label{tab:dummy_stat}
	\begin{tabular}{@{}lccc}
		\toprule
		\multicolumn{4}{c}{Method: cINN}\\ 
		Parameter & Median &1-$\sigma$ &2-$\sigma$ \\
		\midrule
		$x_1$ & 0.51 &[0.36, 0.68] & [0.29, 0.81] \\
		$x_2$ & 0.38 & [0.30, 0.45] & [0.23, 0.50] \\
		$y_1$ & 0.38 & [0.30, 0.45] & [0.23, 0.50] \\
		$y_2$ & 1.28 & [1.20, 1.35] & [1.11, 1.39] \\
		\midrule         
		\multicolumn{4}{c}{Method: MCMC}\\ 
		Parameter & Median &1-$\sigma$ &2-$\sigma$ \\
		\midrule
		$x_1$ & 0.51 &[0.36, 0.68] & [0.29, 0.82] \\
		$x_2$ & 0.38 & [0.30, 0.45] & [0.23, 0.50] \\
		$y_1$ & 0.38 & [0.30, 0.45] & [0.23, 0.50] \\
		$y_2$ & 1.28 & [1.19, 1.35] & [1.11, 1.39] \\
		\midrule         
		\multicolumn{4}{c}{Method: Analytical}\\ 
		Parameter & Median &1-$\sigma$ &2-$\sigma$ \\
		\midrule
		$x_1$ & 0.51 & [0.37,0.69] & [0.29,0.82] \\
		$x_2$ & 0.38 & [0.30,0.45] & [0.23, 0.50] \\
		$y_1$ & 0.38 & [0.30, 0.45] & [0.23, 0.50] \\
		$y_2$ & 1.28 & [1.20,1.36] & [1.11, 1.39] \\
		\bottomrule    
	\end{tabular}
\end{table}

\begin{table}
    \centering
    \caption{Hellinger distances comparing the marginalized posterior distributions for the toy model.}
    \label{tab:dummy_hellinger}
    \begin{tabular}{@{}lcc}
         \hline
         \hline
         Parameter & $h(p_\text{analytical},p_\text{cINN})$ &$h(p_\text{analytical},p_\text{MCMC})$ \\
         \midrule
         $x_1$ & 0.027 & 0.006\\
         $x_2$ & 0.023 & 0.007 \\
         $y_1$ & 0.024 & 0.007 \\
         $y_2$ & 0.047 & 0.006 \\
        \bottomrule         
    \end{tabular}
\end{table}

\section{Results}\label{Sec:Results}
\subsection{Training Performance}\label{Sec:training_perf}
\begin{figure*}
    \centering
    \resizebox{\hsize}{!}{\includegraphics{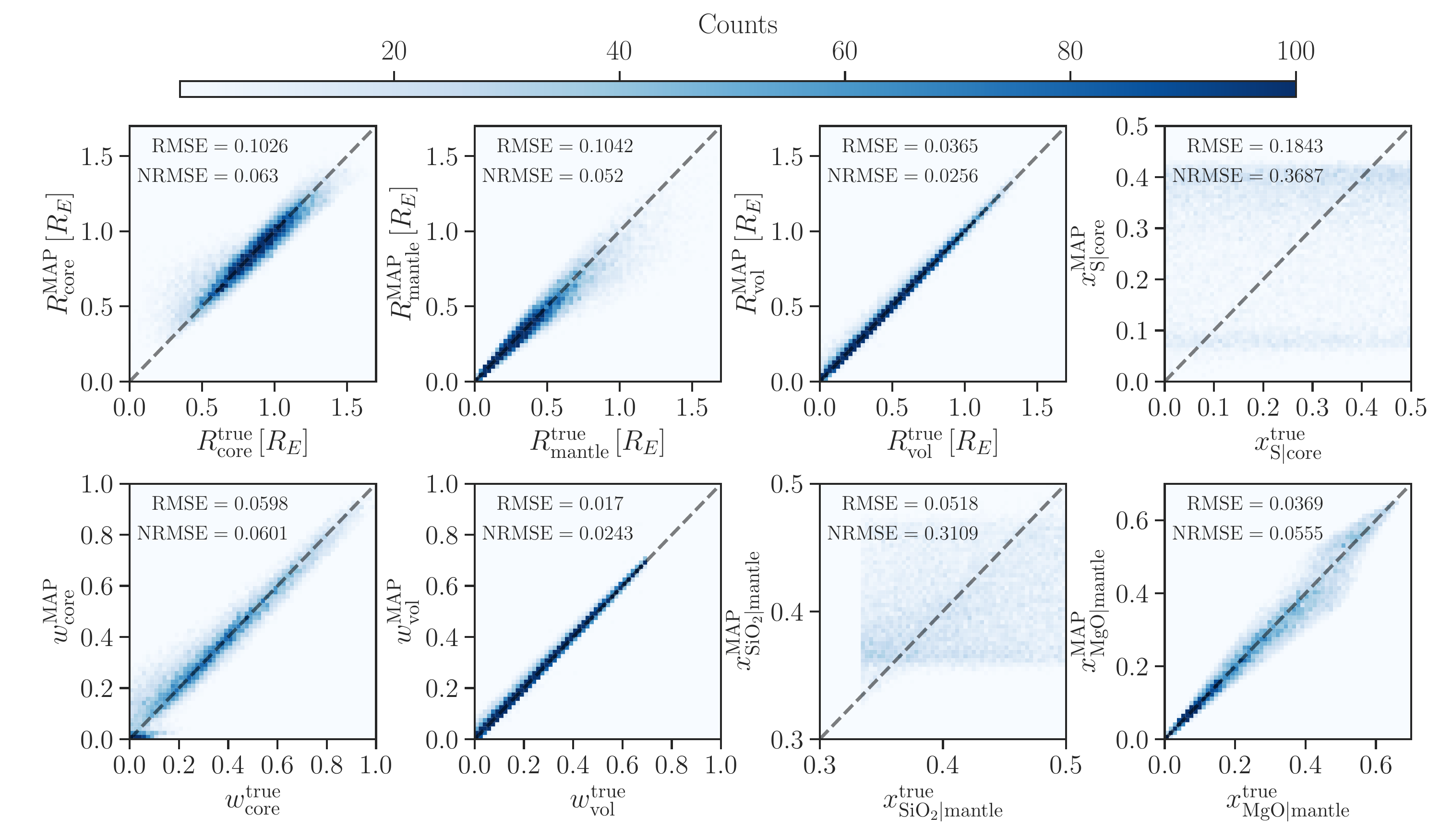}}
    \caption{Distribution of N=20'000, maximum a posteriori probability (MAP) estimates of the trained cINN plotted against the ground truth from the training data set, shown for the model input parameters.}
    \label{fig:true_map}
\end{figure*}

\begin{table}
	\centering
	\caption[Overview of test performance of the cINN.]{Overview of cINN test performance for the planet characterization task.}
	\label{Tab:cinn_performance}
	\begin{tabular}{@{}lcccc}
		\toprule
        Parameter & RMSE & NRMSE & $s$ & $u_{68}$ \\
        \midrule
        $R_\mathrm{core}$ & 0.1026 & 0.0630 & 0.005 & 0.160 \\
        $R_\mathrm{mantle}$ & 0.1042 & 0.0520 & 0.007 & 0.133 \\
        $R_\mathrm{vol}$ & 0.0365 & 0.0256 & 0.002 & 0.043 \\
        $w_\mathrm{core}$ & 0.0598 & 0.0601 & 0.005 & 0.095 \\
        $w_\mathrm{vol}$ & 0.0170 & 0.0243 & 0.002 & 0.022 \\
        $x_\mathrm{SiO_2}|_\mathrm{mantle}$ & 0.0518 & 0.3109 & 0.001 & 0.107\\
        $x_\text{MgO}|_\text{mantle}$ & 0.0369 & 0.0555 & 0.001 & 0.063 \\
        $x_\mathrm{S}|_\mathrm{core}$ & 0.1843 & 0.3687 & 0.002 & 0.340 \\
		\bottomrule 
	\end{tabular}
\end{table}

For the planet characterization task we train a cINN to predict the physical parameters $R_\text{core}$, $R_\text{mantle}$, $R_\text{vol}$ (i.e.~the radii of the core, mantle and surface layer, respectively), $w_\text{core}$, 
$w_\text{vol}$, $x_\text{SiO$_2$}|_\text{mantle}$, $x_\text{MgO}|_\text{mantle}$ and $x_\text{S}|_\text{core}$ from the observables $M_\text{tot}$, $R_\text{tot}$, $x_\text{Si}/x_\text{Fe}|_\text{Planet}$ and $x_\text{Mg}/x_\text{Fe}|_\text{Planet}$, using the database described in Section \ref{Sec:app_training}. Table~\ref{Tab:cinn_performance} summarizes the performance of the trained cINN model on the synthetic, held-out test data for all the target parameters, listing the RMSE and NRMSE for the MAP point estimates, as well as the calibration errors $s$ and median uncertainty at 68\% confidence $u_{68}$ (i.e.~the width of the 68\% confidence interval) for the posterior distributions. Figure~\ref{fig:true_map} also provides 2D histograms comparing the MAP estimates against the corresponding ground truth values. Note that the performance on the synthetic test set is evaluated without error re-sampling, i.e.~assuming perfect observations. For a discussion of the computational cost of the cINN training see Section~\ref{Sec:CompCost}. 

As the histograms and the RMSEs/NRMSEs demonstrate the cINN can recover $R_\text{core}$, $R_\text{mantle}$, $R_\text{vol}$, $w_\text{core}$, 
$w_\text{vol}$ and $x_\text{MgO}|_\text{mantle}$ well with MAP estimates that fall very close or directly on to the ideal 1-to-1 correlation in comparison to the ground truth. For $x_\text{S}|_\text{core}$ and $x_\text{SiO$_2$}|_\text{mantle}$, however, we find that the MAP point estimates cannot match the ground truth values at all with results that are scattered across the entire parameter ranges with no discernible overdensity at the 1-to-1 correlation. Looking at the median calibration errors $s$ of the underlying predicted posterior distributions, we find that the cINN finds very well calibrated solutions (i.e.~posteriors that are neither too broad or too narrow) with values below $0.7$ per cent for all target parameters, including $x_\text{S}|_\text{core}$ and $x_\text{SiO$_2$}|_\text{mantle}$. From the median widths of the 68\% confidence intervals, which are on average on the order of $\approx 0.1$, we find, however, that the posterior distributions tend to be rather broad in general (taking the target parameter ranges into account). 

Looking at the posterior distributions themselves, the issues with the $x_\text{S}|_\text{core}$ and $ x_\text{SiO$_2$}|_\text{mantle}$ MAP estimates result from the fact that the cINN consistently predicts almost perfectly uniform distributions across the parameter ranges for these two parameters for all examples in the test set. In this case performing an MAP estimate simply becomes unfeasible as it merely picks up on minor random fluctuations in these almost uniform distribution rather than identifying distinct peaks in the posteriors. As we show later in our direct comparison of the cINN and an MCMC approach in Figure~\ref{Fig:inn_vs_mcmc1} these almost uniform posterior distributions of $ x_\text{S}|_\text{core}$ and $ x_\text{SiO$_2$}|_\text{mantle}$ are not a flaw of our cINN model, but are also recovered by the MCMC. Given that both cINN and MCMC, thus, return rather uninformative posterior distributions for $x_\text{S}|_\text{core}$ and $x_\text{SiO$_2$}|_\text{mantle}$, we have to conclude that these two physical parameters cannot be constrained from the observables $M_\mathrm{tot}$, $R_\mathrm{tot}$, $x_\text{Si}/x_\text{Fe}|_\text{Planet}$ and $x_\text{Mg}/x_\text{Fe}|_\text{Planet}$. Nevertheless for the remaining physical parameters, the cINN has demonstrated a highly  satisfactory predictive performance on the synthetic test data.

\subsection{Comparing the cINN to an MCMC method, in the case of K2-111 b}
\begin{table}
    \centering
    \caption{Benchmark case to compare the cINN with a regular MCMC sampler.}
    \label{tab:cases_ov}
    \begin{tabular}{@{}lllll}
         \hline
         \hline
         Planet & Mass [M$_\text{E}$] & Radius [R$_\text{E}$] & $\frac{x_\text{Si}}{x_\text{Fe}}|_\text{Planet}$ & $\frac{x_\text{Mg}}{x_\text{Fe}}|_\text{Planet}$\\
         \midrule
         K2-111 b$^{(1)}$ & $5.29^{+0.76}_{-0.77}$ &$1.82^{+0.11}_{-0.09}$ &$1.82^{+0.48}_{-0.38}$ & $2.51^{+0.85}_{-0.63}$\\
        \bottomrule         
    \end{tabular}
    \tablebib{
    (1): \citet{mortier_k2-111_2020}
    }
\end{table}

In order to demonstrate that our cINN provides accurate posterior distributions of planetary parameters, we considered the case of K2-111 b \citep{mortier_k2-111_2020}, which was observed to be a planet of radius $1.82_{-0.09}^{+0.11}$  $R_\text{E}$ with a mass of $5.29 ^{+0.76}_{-0.77}$ $M_\text{E}$. Its high mean density of $4.81_{-1.01}^{+1.25}$ g/cm$^3$ implies that the planet is very likely to have only a tiny gas envelope, making it a well suited example for our purpose. We also assume that the composition of the photosphere of of K2-111 is identical to the one of the planet in terms of the elemental ratios of the refractory elements Mg, Fe and Si. The elemental ratios given in \citet{mortier_k2-111_2020} are $1.82^{+0.48}_{-0.38}$ for Si/Fe and $2.51^{+0.85}_{-0.63}$ for Mg/Fe. With this observation we sample the posterior distribution of the forward model parameters using the cINN method as described in Sect. \ref{Sec:posterior_with_uncert}.

\subsubsection{MCMC setup}
For this comparison we also sample the posterior probability distribution using a MCMC method as it is often employed to infer planetary interiors \citep[see e.g.,][]{haldemann_mcmc_2021,dorn_bayesian_2017}.
In particular we use the adaptive Metropolis-Hastings MCMC algorithm \citep{haario_adaptive_2001}, sampling the posterior distribution of 
\begin{equation}
    p(\textbf{x}|\textbf{y}_\text{obs}) \propto \mathbb{L}(\textbf{y}_\text{obs}|\textbf{x})\, \mathbb{P}(\textbf{x})
\end{equation}
given the same forward model $f(\cdot)$ as described in Sect. \ref{Sec:forward}. We consider the same prior $\mathbb{P}(x)$ as when constructing the training data (see Tab. \ref{tab:input_dist}).  The likelihood $\mathbb{L}(\textbf{y}_\text{obs}|\textbf{x})$ is calculated using
\begin{equation}
    \mathbb{L}(\textbf{y}_\text{obs}|\textbf{x}) = \frac{1}{(2\pi)^{N/2}\left(\prod_{i=1}^{N}\sigma_i\right)^{1/2}}\exp\left(-\frac{1}{2}\sum^N_{i=1}\frac{(f(\textbf{x})_i-\mu_i)^2}{\sigma_i^2}\right),
\end{equation}
where $\mu$ and $\sigma^2$ are the mean and variance of $\textbf{y}_\text{obs}$ and N is the number of dimensions or parameters of $\textbf{y}_\text{obs}$.

\begin{figure}
    \centering
	\resizebox{\hsize}{!}{\includegraphics{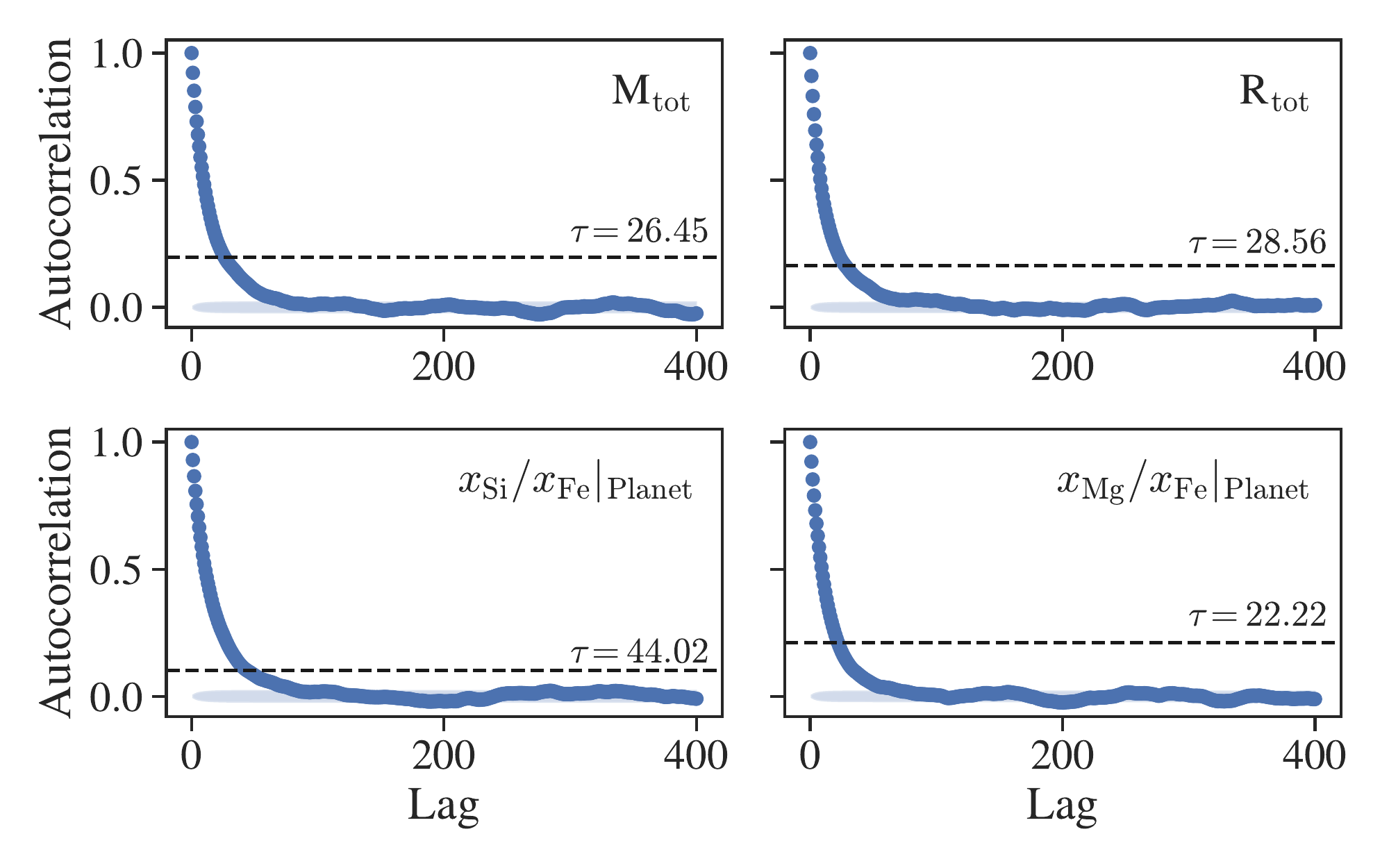}}
	\caption{Autocorrelation as a function of lag calculated from the Markov chain of the four output parameters of the forward model. The autocorrelation time $\tau$ of each parameter was calculated following \cite{hogg_data_2018}. The dashed lines indicate the autocorrelation when the lag is equal to the autocorrelation time.}
	\label{Fig:acf_mcmc}
\end{figure}

The MCMC was initialized at a random location in the sample space. We then run the MCMC for $\sim$\,$5\cdot10^5$ steps. The resulting Markov chain had an autocorrelation time $\tau$ between 28 and 44 steps for the four output parameters (see Fig. \ref{Fig:acf_mcmc}). The autocorrelation time was calculated following \citet{hogg_data_2018}.
To generate independent samples from the Markov chain we begin by discarding the first 2000 steps along the Markov chain due to burn in and then, accounting for the maximum autocorrelation time, add every 50th step along the chain to the set of independent samples. This resulted in a total of  $10^4$ independent samples from the Markov chain, which is sufficient given the shape and number of dimensions of the posterior probability distribution.

\subsubsection{Comparison of the marginalized posterior distributions}
In order to compare the performance of the cINN with the MCMC method, we show in Table \ref{Tab:mcmc_vs_cinn} a summary of the key statistics of the 1D-marginalized posterior PDF.
For each model parameter the median as well as the centered 1-$\sigma$ and 2-$\sigma$ intervals (containing 68\% respectively 95\% of all samples) are shown.
Looking at the median values one can see that both methods return almost identical results with a maximum difference of $\sim 5.8\%$. For the boundaries of the centered 1-$\sigma$ and 2-$\sigma$ regions the differences are larger, notably for the lower boundaries of $R_\text{vol}$, $w_\text{vol}$, $x_\mathrm{S}|_\mathrm{core}$, $x_\text{FeO}|_\mathrm{mantle}$ and $x_\text{SiO$_2$}|_\mathrm{mantle}$.
However, the Hellinger distances $h$ between the marginalized posteriors predicted by the two methods are very low ($h<0.05$) for all parameters except $R_\text{vol}$, $w_\text{vol}$ and $R_\text{tot}$ (see Table~\ref{Tab:mcmc_vs_cinn_hellinger} ). The largest Hellinger distance of 0.103 for $R_\text{vol}$ is equivalent to the distance of two standard normal distributions whose median is shifted by 0.25.

\begin{figure*}
    \centering
    \resizebox{0.45\hsize}{!}{\includegraphics{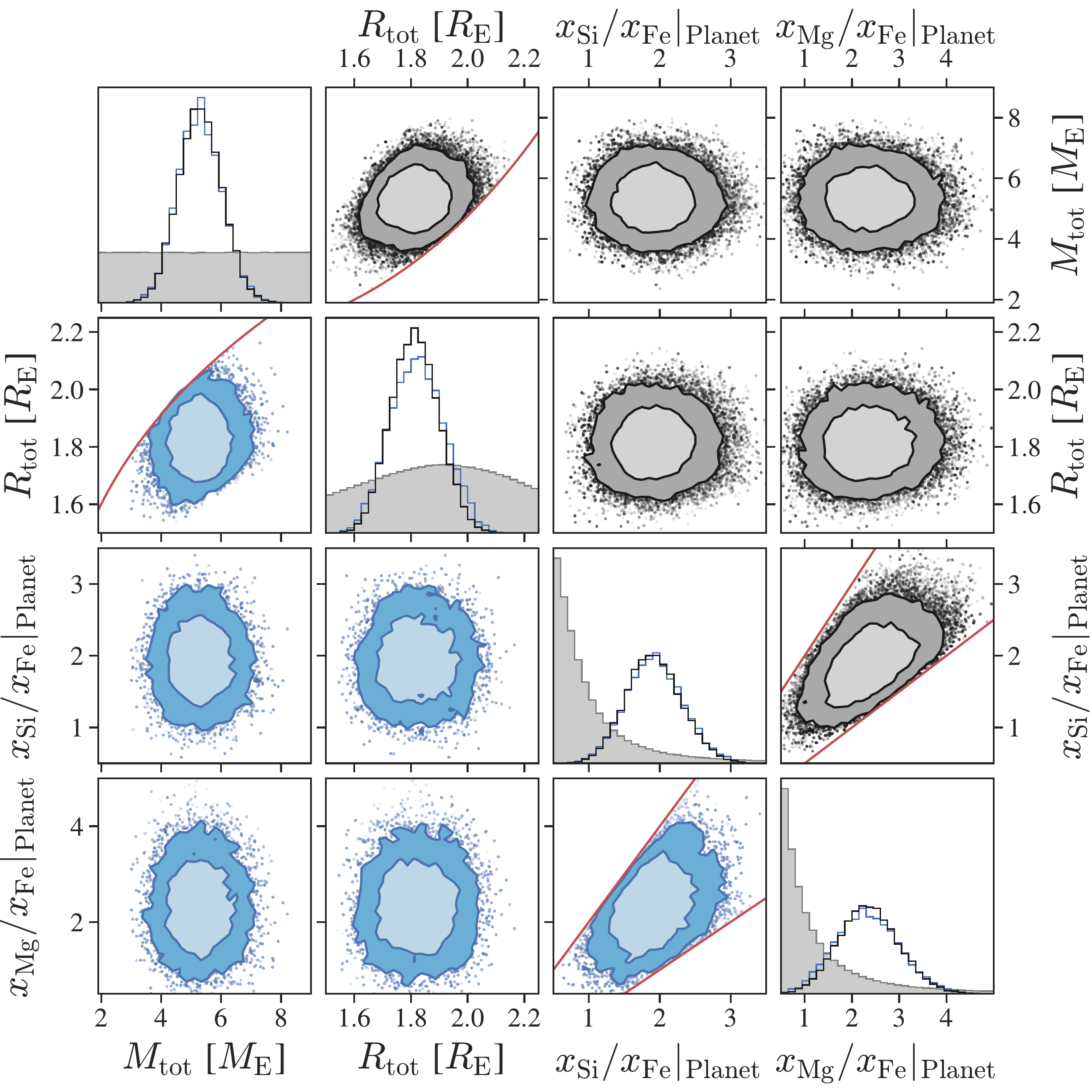}}
    \resizebox{0.8\hsize}{!}{\includegraphics{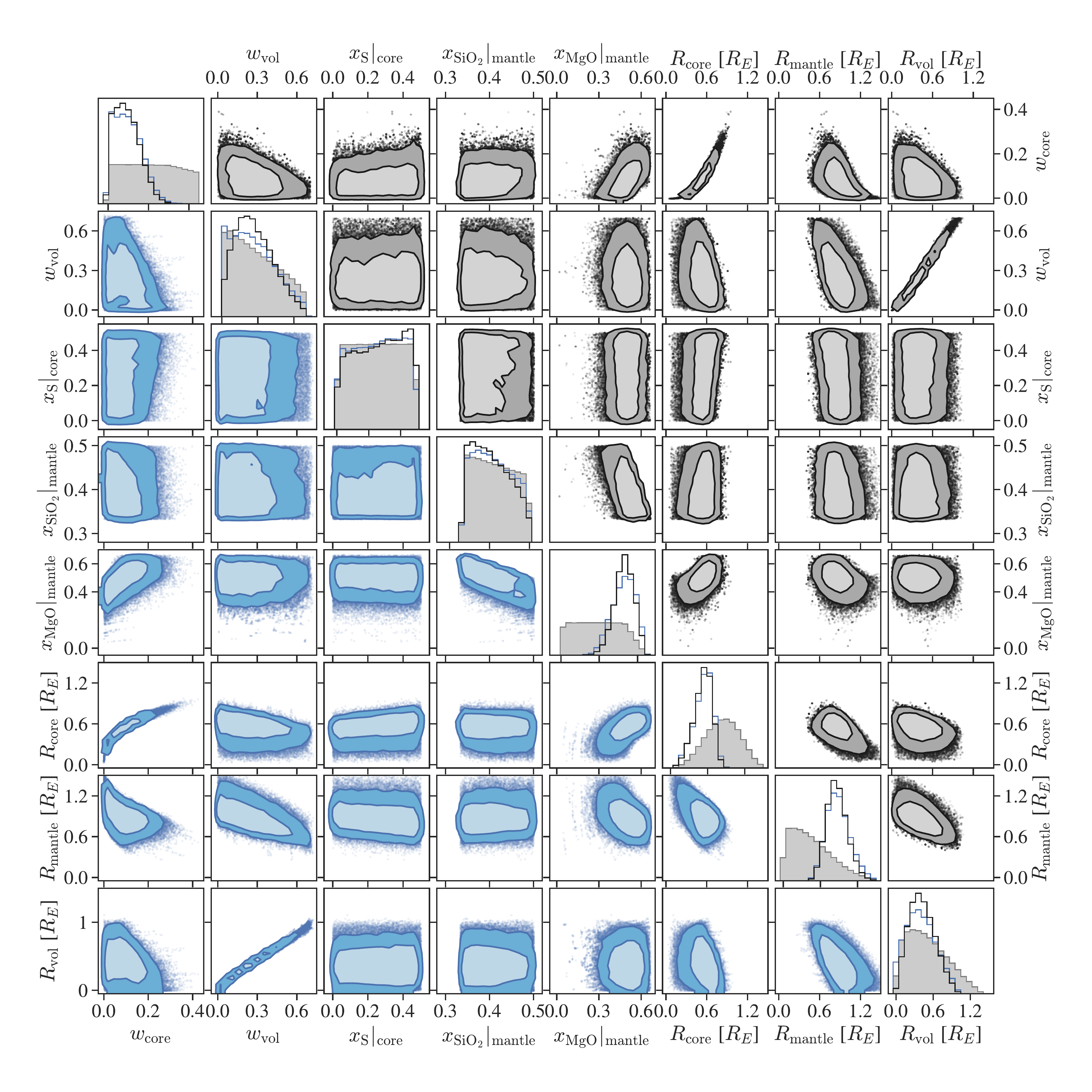}}
	
	\caption{Comparison of the cINN and an MCMC method when applied to K2-111 b. The data in the lower triangle of each subfigure (blue points) was generated with the cINN method, while the data in the upper triangle (black points) was generated with the MCMC sampler. In the diagonal panels also the marginalized prior probability (grey) is shown.  The light shaded areas in the 2D diagrams indicate the 68\%-HDR while the dark shaded areas are the 89\%-HDR. The red solid line indicates the limits of the forward model as discussed in Sec.~\ref{Sec:limits}.}
	\label{Fig:inn_vs_mcmc1}
\end{figure*}

In Fig.~\ref{Fig:inn_vs_mcmc1} 
we show the pairwise marginalized 2D posterior PDF of all  parameters. On the diagonal of these figures we also show the 1D histograms of the marginalized PDF as well as the prior distribution of the points in the training data.
Overall the shape of the pairwise marginalized 2D posterior PDF is very similar between the two methods. One can also see in Fig.~\ref{Fig:inn_vs_mcmc1} that the cINN method can be used close to the boundary of the training data (red lines).
For the 1D histograms the biggest difference is again seen for $R_\text{vol}$ and $w_\text{vol}$, especially towards dry compositions.


\begin{figure}
	\centering
	\resizebox{\hsize}{!}{\includegraphics{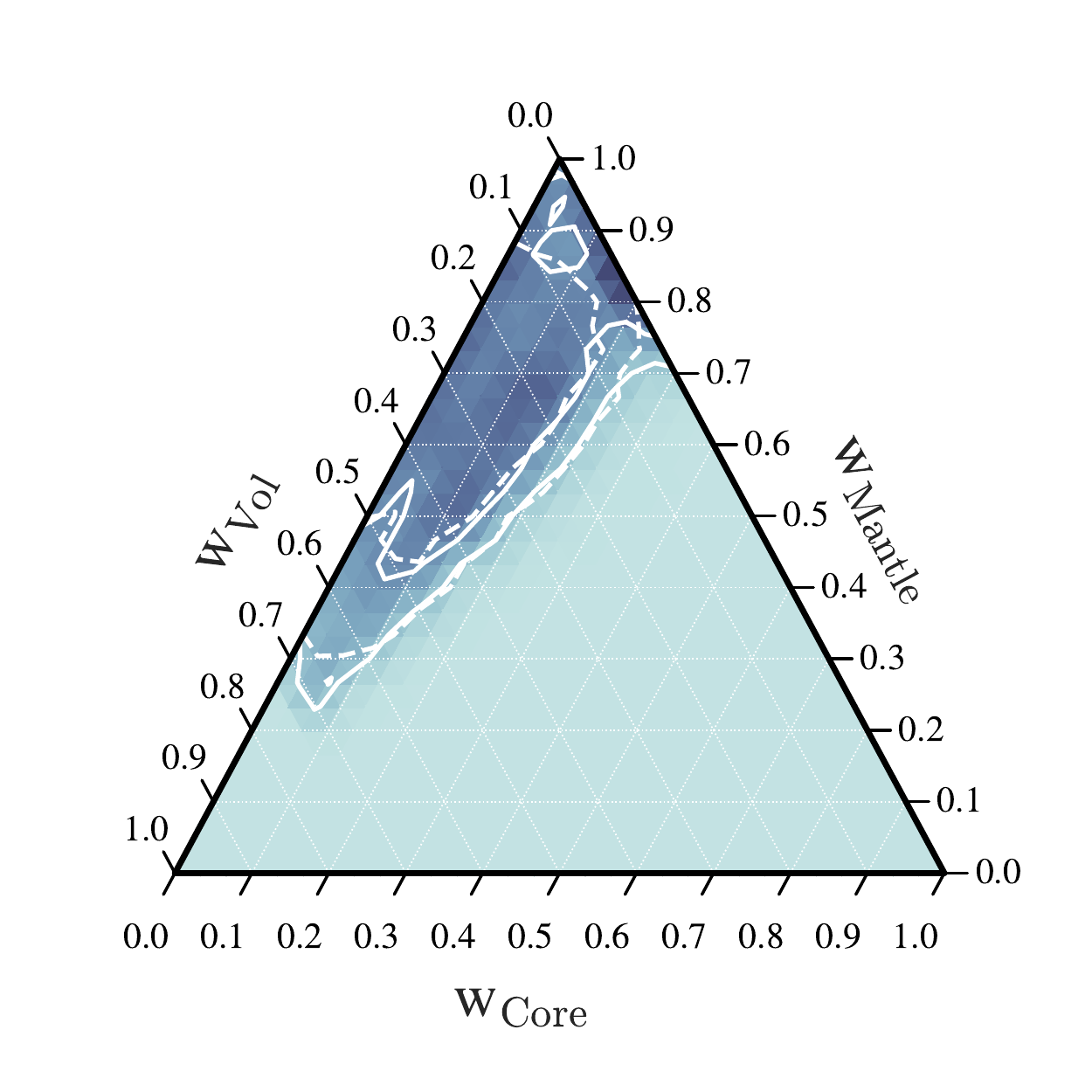}}
	\resizebox{\hsize}{!}{\includegraphics{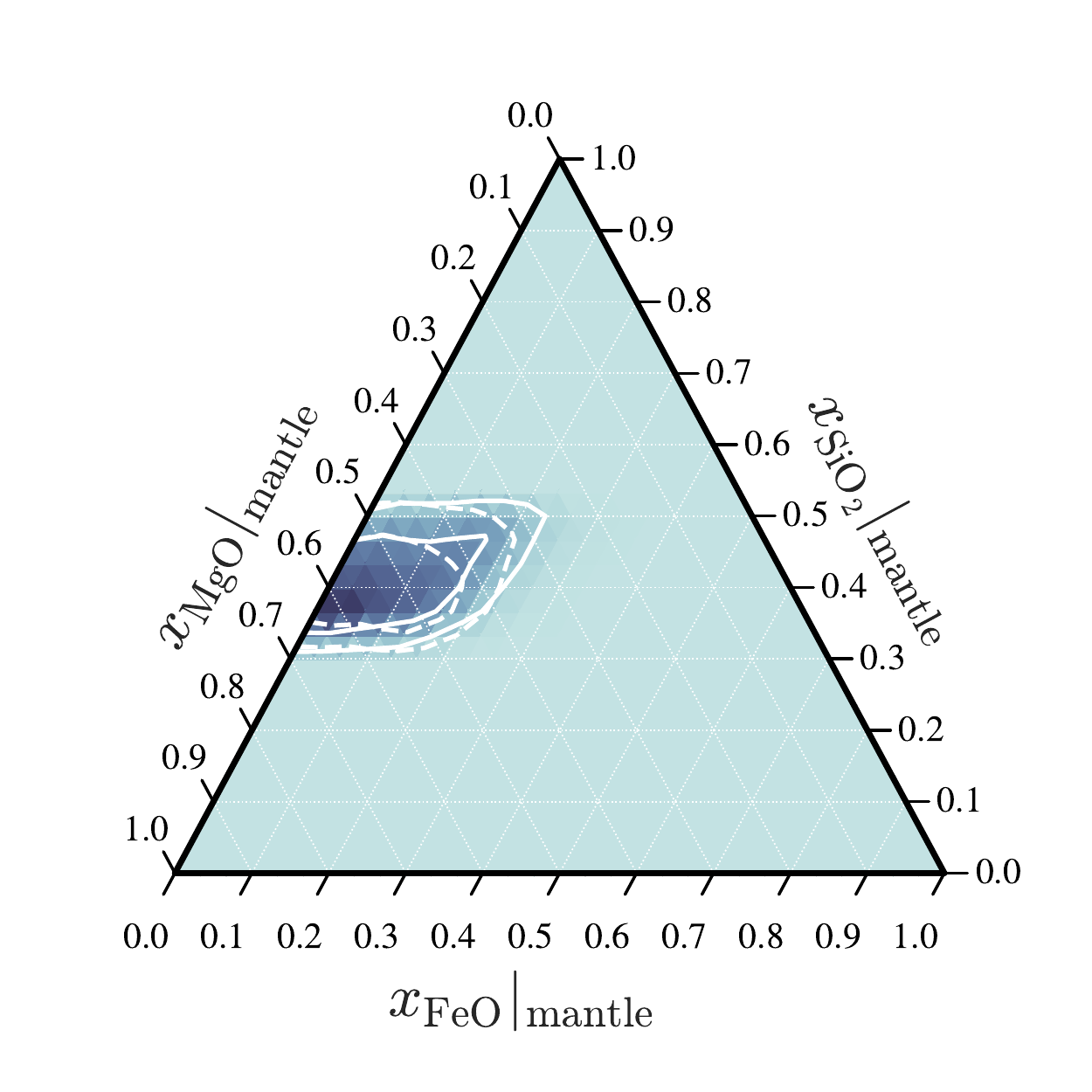}}

	\caption{\textit{Top panel}: Kernel density estimate of the layer mass fractions as determined with the cINN and plotted on a ternary diagram. \textit{Bottom panel:}  Kernel density estimate of the mantle composition. The kernel density in both panels was estimated using a Gaussian Kernel with standard deviation of 0.2. The white lines indicate the contours of the 68\%-HDR and 95\%-HDR. For comparison also the HDR from the posterior calculated with the MCMC are shown (dashed lines). }
	\label{Fig:cinn_ternary2}
\end{figure}

The layer mass fractions and the mantle composition are each a set of compositional variables\footnote{Hence they sum up to one.}, thus in Fig.~\ref{Fig:cinn_ternary2} their distribution on the ternary diagram is shown. For the layer mass fractions, one can see a very good agreement in the region above 0.1 $w_\text{vol}$. But as it is also present in the 1D marginalized distributions, there are fewer samples with low volatile content in the Markov chain than in the set generated using the cINN. For the mantle composition the posterior distribution of both methods is centred around compositions with 60\% MgO and less than 10\% FeO. Compared to the MCMC method the cINN predicts slightly more compositions with larger amounts of SiO$_2$ than MgO, which results in the two kinks in the contours on the ternary diagram.

\subsubsection{The recalculation error of the trained cINN for K2-111~b}
\begin{figure}
	\centering
		\resizebox{\hsize}{!}{\includegraphics{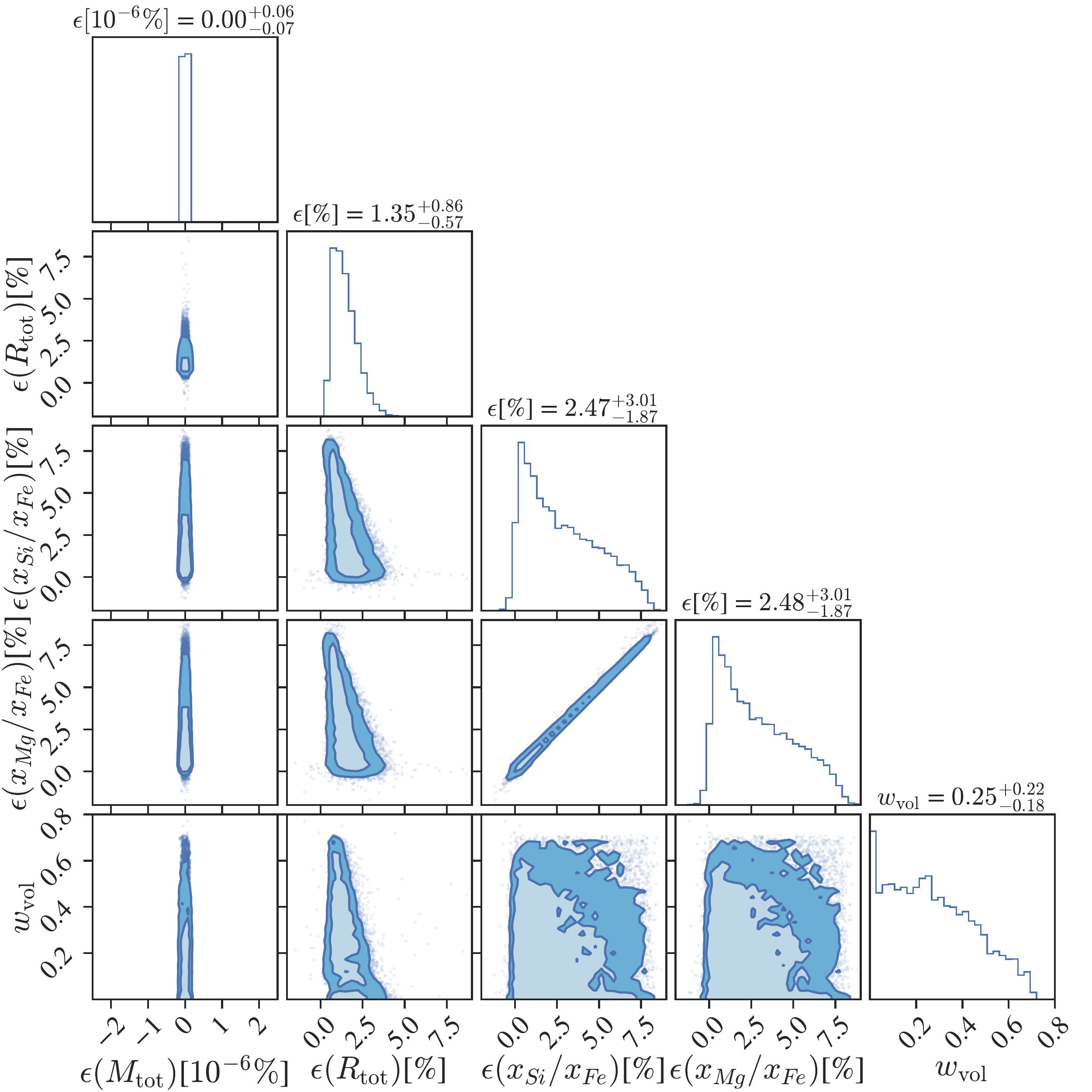}}	
	\caption{Recalculation errors of the model output variables for the set of samples generated with the cINN for the case of K2-111 b.}
	\label{Fig:cinn_recalc}
\end{figure}

To assess the quality of the inverse mapping from $\textbf{y}$ to $\textbf{x}$, we calculated for each sample $\textbf{x}^{(i)}$ generated with the cINN the recalculation error $\epsilon$. For that we apply to each sample $\textbf{x}^{(i)}$ again the forward model $f(\cdot)$. The recalculation error is then given by the relative difference between $f(\textbf{x}^{(i)})$, and the $\textbf{y}^{(i)}$ given to the cINN to generate $\textbf{x}^{(i)}$ initially. $\epsilon$ is calculated for each component $k$ of $\textbf{y}^{(i)}$, i.e., 
\begin{equation}
	\epsilon(k) = 100\cdot\frac{f(g(\textbf{z},\textbf{c}=\textbf{y}^{(i)}))_k - y_k^{(i)}}{y_k^{(i)}} = 100\cdot\frac{f(\textbf{x}^{(i)})_k - y_k^{(i)}}{y_k^{(i)}}.\label{Eq:recalc_error}
\end{equation}
In Figure~\ref{Fig:cinn_recalc} we show the distribution of the recalculation errors of all data variables together with $w_\text{vol}$, given the set of $\textbf{x}^{(i)}$ generated for the case of K2-111 b.
One can see that the cINN learned best the mapping of the total mass, while the median recalculation error of the other variables is between 1.3\% and 2.5\%. Figure \ref{Fig:cinn_recalc} also indicates that the recalculation errors are increasing towards low values of $w_\text{vol}$, especially below 0.1. This indicates that the quality of the cINN mapping in this region is not yet optimal, which likely explains the observed differences in the posterior distribution of $w_\text{vol}$ and $R_\text{vol}$ between the MCMC and the cINN method.

\begingroup
\setlength{\tabcolsep}{4pt}
\begin{table*}
	\centering
	\caption[Various statistics of the marginalized posterior distribution of the forward model parameters.]{Various statistics of the marginalized posterior distribution of the forward model parameters. Comparing the cINN method with a Metropolis-Hastings MCMC scheme when applied to K2-111 b.}
	\label{Tab:mcmc_vs_cinn}
	\begin{tabular}{@{}lccccccccc}
		\toprule
		\multicolumn{4}{c}{Method: cINN}&\multicolumn{3}{c}{Method: MCMC}&\multicolumn{3}{c}{Difference: $100\cdot(x_\text{cINN}-x_\text{MCMC})/x_\text{MCMC}$}\\
		Parameter & Median &1-$\sigma$ &2-$\sigma$& Median &1-$\sigma$ &2-$\sigma$& $\Delta$Median &$\Delta$1-$\sigma$ &$\Delta$2-$\sigma$\\
		\midrule
		$w_\mathrm{core}$ & 0.09 & [0.03, 0.17] & [0.00, 0.24]& 0.09 & [0.03, 0.16] & [0.01, 0.23]& 5.84 & [-4.30, 6.60] & [-23.19, 6.95] \\
		$w_\mathrm{rock}$ & 0.64 & [0.45, 0.81] & [0.30, 0.92]& 0.64 & [0.47, 0.78] & [0.33, 0.89]& 0.69 & [-4.01, 3.51] & [-8.66, 3.32] \\
		$w_\mathrm{vol}$ & 0.25 & [0.07, 0.46] & [0.00, 0.64]& 0.26 & [0.11, 0.44] & [0.03, 0.61]& -3.85 & [-34.35, 4.37] & [-89.56, 4.35] \\
		$x_\mathrm{S}|_\mathrm{core}$ & 0.26 & [0.08, 0.42] & [0.01, 0.48]& 0.27 & [0.09, 0.43] & [0.01, 0.49]& -5.10 & [-8.83, -3.23] & [-16.19, -1.32] \\
		$x_\text{FeO}|_\mathrm{mantle}$ & 0.09 & [0.03, 0.17] & [0.01, 0.24]& 0.10 & [0.03, 0.17] & [0.00, 0.23]& -4.56 & [-8.95, 3.25] & [24.75, 5.10] \\
		$x_\text{SiO$_2$}|_\mathrm{mantle}$ & 0.41 & [0.36, 0.47] & [0.34, 0.49]& 0.40 & [0.35, 0.46] & [0.34, 0.49]& 1.47 & [0.79, 1.49] & [0.14, 0.33] \\
		$x_\text{MgO}|_\mathrm{mantle}$ & 0.50 & [0.39, 0.58] & [0.30, 0.63]& 0.50 & [0.42, 0.57] & [0.33, 0.63]& -0.19 & [-6.02, 1.89] & [-8.79, 1.26] \\
		$R_\mathrm{core}$ [$R_E$] & 0.55 & [0.37, 0.68] & [0.19, 0.78]& 0.54 & [0.37, 0.67] & [0.21, 0.76]& 2.22 & [-0.51, 2.10] & [-7.53, 2.08] \\
		$R_\mathrm{rock}$ [$R_E$] & 0.88 & [0.70, 1.08] & [0.56, 1.31]& 0.86 & [0.71, 1.04] & [0.57, 1.23]& 1.89 & [-0.39, 4.67] & [-2.29, 6.33] \\
    $R_\mathrm{vol}$ [$R_E$]& 0.38 & [0.15, 0.66] & [0.01, 0.88]& 0.40 & [0.21, 0.64] & [0.07, 0.85]& -4.89 & [-29.47, 2.84] & [-85.79, 4.07] \\
		\bottomrule 
	\end{tabular}
	\tablefoot{
	$\Delta 1$-$\sigma$ and $\Delta2$-$\sigma$ are the relative differences of the interval boundaries of the $1$-$\sigma$ and $2$-$\sigma$ intervals. The $w_\mathrm{rock}$ and $x_\text{FeO}|_\mathrm{mantle}$ were calculated from the other layer mass fractions, respectively, other mantle oxide fractions.
	}
\end{table*}
\endgroup

\begin{table}
	\centering
	\caption{Hellinger distance metric between the marginalized posterior distributions of the cINN and MCMC method.}
	\label{Tab:mcmc_vs_cinn_hellinger}
	\begin{tabular}{@{}lc}
		\toprule
		Parameter & Hellinger distance $h$ \\
		\midrule
		$M_\textbf{}{tot}$ & 0.027  \\
		$R_\textbf{}{tot}$ & 0.067  \\
		$x_\text{Si}/x_\text{Fe}|_\text{Planet}$ & 0.028 \\
		$x_\text{Mg}/x_\text{Fe}|_\text{Planet}$ & 0.041 \\
		$w_\text{core}$ & 0.047  \\
		$w_\text{vol}$ & 0.098  \\
		$x_\text{S}|_\text{core}$ & 0.037 \\
		$x_\text{SiO$_2$}|_\text{mantle}$ & 0.039 \\
		$x_\text{MgO}|_\text{mantle}$ & 0.089 \\
		$R_\text{core}$ & 0.043\\
        $R_\text{rock}$ &0.062\\
        $R_\text{vol}$&0.103\\
		\bottomrule 
	\end{tabular}
\end{table}

\section{Discussion}\label{Sec:Discussion}
\subsection{Comparison of the computational cost}\label{Sec:CompCost}
One major motivation for using cINNs to infer planetary compositions is, to reduce the time needed to perform a single inference. We give in the following an overview over the encountered computational cost when using the two methods shown in this work, i.e., an adaptive Metropolis Hastings MCMC method and the cINN method. 

The MCMC method is calculated in a sequential manner, i.e., for each step of the Markov chain a forward model is run until a sufficient number of steps are generated.
Hence its computational cost scales linearly with the number of steps of the generated Markov chain.
For the used forward model, it was proven sufficient to generate on the order of $5\cdot 10^5$ forward models.
On a single core of an Intel Xeon Gold 6132 processor running at 2.6 GHz one forward model takes between one and two seconds to compute.
Therefore performing one planetary structure inference takes in total approximately 8.7 days to compute on a single core.

Instead of computing a single Markov chain it is also possible to initialize multiple chains in parallel or use an ensemble method such as in \textit{emcee} \citep{foreman-mackey_emcee_2013}. Such a parallelized approach allows to leverage the availability of multi-core CPU architectures, though for efficient tuning of the proposal distribution one needs a minimal length of the Markov chain on the order of $\sim 10^4$ samples. Using 28 cores of an Intel Xeon Gold 6132 CPU one has to compute $10^4$ steps for each of the 28 Markov chains to account for burn-in and tuning of the proposal distribution. This results in a total of $7.8\cdot10^5$ forward models to be calculated by 28 parallel MCMC chains, which takes approximately 12 hours.

The total computational cost of the cINN method in contrary is split into four parts. The computation time needed to generate the training data, the time needed to train the cINN, the hyperparameter search for optimal training (i.e., determining learning rate, network architecture, etc.) and the time needed to sample the posterior using the cINN. 

For this project the training data was generated calculating $5.9 \cdot 10^6$ forward models.
With an average run time of 1.5 s this would take 102 days to run on a single core, whereas using a compute node with 28 cores, the data set can be generated within 3.7 days. 
With the training data at hand training the cINN itself takes between 2 and 3 hours on a single GPU \footnote{The training and inference were performed on a compute node at the \textit{Interdisziplinären Zentrum für Wissenschaftliches Rechnen} (IWR) in Heidelberg, which consists of 2 x 14-Core Intel Xeon Gold 6132 @ 2.6 GHz and 10 x Nvidia Titan Xp @ 1.6 GHz, but only one GPU was used.}.
Then a hyperparameter search is necessary to find the parameters for optimal training (i.e., learning rate, number of the neural network layers, network layer widths, etc.).
For this study we performed 23 trials to find the optimal parameters, thus repeating the training of the cINN 23 times. 
A single inference of an exoplanets composition using the trained cINN, on the same GPU as mentioned above, can be performed in 5 minutes.

When comparing the computational cost of the two methods it is clear that for a single inference, the MCMC method is far cheaper given the large number of training data needed to train the cINN.
But when multiple inferences using the same forward model should be performed, then generating the training data will contribute less to the total computational cost the more inferences are performed.
Taking into account that the hyper-parameter search does in both cases not need to be repeated (except if the MCMC method is used for very different data) then the cINN already becomes the more efficient method if the same forward model is used for more than 10 planetary structure inferences.
In Table \ref{Tab:cinn_timing} we show a summary of the computational cost of the two methods.
\footnote{Note that so far we used a forward model without an additional atmosphere layer.
Including an atmosphere in the forward model would add another 2-3 more input parameters and also make the forward model computationally more expensive.
From experience with running structure inference models which include atmosphere layers, a number of $5\cdot10^5$-$10^6$ samples would be necessary to be generated for the Markov Chain. We did not yet create a database of forward models including an atmosphere, hence we can not conclude how this change would affect the computational cost.
Though the time needed for inference should remain on the order of minutes for the cINN approach.}

\begin{table}
	\centering
	\caption[Overview over the computational cost for the two inference methods]{Overview over the computational cost for the two inference methods}
	\label{Tab:cinn_timing}
	\begin{tabular}{lcc}
	\toprule
	& \multicolumn{2}{c}{Computing time}\\
	Method & Total & Single inference\\
	\midrule
	cINN$^\ddag$ & 105 days & 5 min\\
	MCMC$^\dagger$ & 8.7 days & 8.7 days \\
	parallel MCMC$^\dagger$ & 13.5 days & 12 hours\\
	\bottomrule
	\end{tabular}
    \tablefoot{
	\tablefoottext{$\dagger$}{Run on a compute node containing 2 x 14-Core Intel Xeon Gold 6132 CPUs @ 2.6 GHz.}
	\tablefoottext{$\ddag$}{Run on a compute node containing 2 x 14-Core Intel Xeon Gold 6132 CPUs @ 2.6 GHz and 10 x Nvidia Titan Xp GPUs @ 1.6 GHz.}
    }
\end{table}

\subsection{Comparison to the initial characterization of K2-111 b}
The exoplanet K2-111 b was also characterized in \citet{mortier_k2-111_2020}. In their work two different internal structure models were used for characterization, one considering four layers, i.e., iron core, silicate mantle, water layer and H/He envelope and the other model only considering two layers, i.e., an iron core with a surrounding mantle. Using the first model, the inferred bulk composition of K2-111 b was $w_\text{core}=0.10^{+0.07}_{-0.07}$, $w_\text{mantle}=0.68^{+0.13}_{-0.14}$, $w_\text{water}=0.20^{+0.16}_{-0.13}$ and $\log_{10} w_\text{H/He}=-8.76^{+2.20}_{-2.21}$. 
Note that the inferred small amount of H/He by \citet{mortier_k2-111_2020} is one reason we chose this exoplanet for our work, since we trained the cINN so far only on planetary structures without H/He layers.

When we compare this to the results obtained using the cINN, which are given in Table \ref{Tab:mcmc_vs_cinn}, 
we see that the inferred core mass fraction is almost the same, while the mantle mass fraction is slightly smaller and the water mass fraction is slightly larger than in \citet{mortier_k2-111_2020}.
They also inferred a mantle composition of $x_\text{FeO}|_\text{Mantle}=0.09^{+0.07}_{-0.06}$, $x_\text{SiO$_2$}|_\text{Mantle}=0.39^{+0.05}_{-0.04}$ and $x_\text{MgO}|_\text{Mantle}=0.51^{+0.06}_{-0.06}$, together with a core composition of $x_\text{S}|_\text{Core}=0.27^{+0.16}_{-0.18}$.
This is also in good agreement with the prediction results of the cINN. 
Do note, however, that the good agreement in $x_\text{S}|_\text{Core}$ and $x_\text{SiO$_2$}|_\text{Mantle}$ may only be by chance here, as both the cINN and MCMC regard these two parameters as largely unidentifiable from the available observations (see Section \ref{Sec:training_perf}).
At the same time, the inference method used in \citet{mortier_k2-111_2020} can merely constrain $x_S|\text{Core}$ and $x_\text{SiO$_2$}|_\text{Mantle}$ beyond the constraints, given by the used structure model.
 
The reason for the small difference in the mantle mass fraction and water mass fraction,
is likely that they used a four layer model including a H/He layer. Although K2-111 b has a very small H/He content in mass \citep[$\sim 10^{-8}$ $M_\text{E}$ as found in][]{mortier_k2-111_2020}, this small H/He layer can still have a radius contribution of up to $\sim 0.1$ $R_\text{E}$, given the planet's large equilibrium temperature ($T_\text{eq} = 1309 $ K). Thus one can expect our results to slightly differ from their study. 
Taking the difference in model setup into account, we conclude that our results agree well with the characterization performed by \citet{mortier_k2-111_2020}.



\section{Conclusions}\label{Sec:Conclusions}
In this work we discussed how one can use invertible neural networks, in particular the conditional invertible neural network (cINNs), to characterize the interior structure of exoplanets. A task for which mainly Markov Chain Monte Carlo (MCMC) methods were used so far.

Compared to the cINN version initially proposed by \citet{Ardizzone2019b} for point estimates, we show in this work how one can adapt the method when facing noisy data. We validated this approach using a toy model, for which we compared the cINN's performance against a regular Metropolis Hastings MCMC.

Then we applied the method to the exoplanet K2-111~b, inferring its composition.  For that we trained a cINN on a simplified internal structure model for exoplanets and showed, that also in this case, cINNs offer a computationally efficient alternative to the MCMC sampler commonly used for Bayesian inference. 

In the benchmark of K2-111 b only minor differences can be seen between the MCMC methods and the cINN method.
The largest differences 
appeared in the marginalized posterior distribution of $R_\text{vol}$ and $w_\text{vol}$.
Computing the recalculation error of the benchmark case showed that the largest errors in total radius appeared for small values of $w_\text{vol}$. This falls in line with the observed differences in the marginalized posterior distributions of $w_\text{vol}$ and $R_\text{vol}$.
Hence it is likely that the difference between the two methods will become smaller if the training of the cINN can be further improved.
Nevertheless the two methods return very similar posterior distributions of the model parameters.

A key benefit of using cINNs over an MCMC method is the fact that the majority of the computational cost of the method occurs during generation of the training data and training but not during the inference.
This allows to reduce the computational time spent for inference by almost four orders of magnitude compared to a regular MCMC method.
In order to have an overall benefit in computational cost against the MCMC method used in this work, the cINN needs to be used to infer more than approximately 10 planetary structures.

While other authors successfully used neural networks to predict the output of their forward models \citep[e.g.][]{alibert_using_2019,baumeister_machine-learning_2020}, 
this work shows that it is also possible to train a neural network which encapsulates the full inverse problem. 
\begin{acknowledgements}
J.H. and Y.A. acknowledge the support from the Swiss National Science Foundation under grant 200020\_172746. V.K.\ and R.S.K.\ thank for funding from the Deutsche Forschungsgemeinschaft (DFG, German Research Foundation) under Germany’s Excellence Strategy EXC-2181/1 - 390900948 (the Heidelberg STRUCTURES Cluster of Excellence). They  also acknowledge financial support from the European Research Council (ERC) via the ERC Synergy Grant "ECOGAL: Understanding our Galactic ecosystem -- From the disk of the Milky Way to the formation sites of stars and planets" (grant 855130), and they thank for financial support from DFG via the collaborative research center (SFB 881, Project-ID 138713538) ”The Milky Way System” (subprojects A1, B1, B2, and B8). The group makes use of computing resources provided by the state of Baden-W\"{u}rttemberg through bwHPC and the German Research Foundation (DFG) through grant INST 35/1134-1 FUGG. Data are stored at SDS@hd supported by the Ministry of Science, Research and the Arts Baden-Württemberg (MWK) and DFG through grant INST 35/1314-1 FUGG.

\\
\\

\textit{Software.} For this publication the following software packages have been used: 
\href{https://matplotlib.org/}{Python-matplotlib} by \citet{Hunter_2007}, \href{https://seaborn.pydata.org/}{Python-seaborn} by \citet{waskom2020seaborn},
\href{https://corner.readthedocs.io}{Python-corner} by \citet{corner}
\href{https://zenodo.org/record/2628066}{Python-ternary} by \citet{marc_marcharperpython-ternary_2019},
\href{https://numpy.org/}{Python-numpy},
\href{https://pandas.pydata.org/}{Python-pandas}.
The cINN is based on the FrEIA framework available at \url{https://github.com/VLL-HD/FrEIA}.

\end{acknowledgements}

\bibliographystyle{aa} 
\bibliography{references.bib}
\end{document}